\documentclass[10pt,preprint]{aastex}
\shorttitle{Heavy Elements in M31}
\shortauthors{Worthey \& Espa\~na}

\def\farcsec{\hbox{$.\!\!^{\prime\prime}$}}

\begin{document}

\title{M31's Heavy Element Distribution and Outer Disk}

\author{Guy Worthey, Aubrey Espa\~na}
\affil{Department of Physics and Astronomy, Washington State University, Pullman, WA
99164-2814}

\author{Lauren A. MacArthur}
\affil{Department of Physics and Astronomy, University of British
Columbia, Vancouver, BC V6T 1Z1}

\author{St{\'e}phane Courteau}
\affil{Department of Physics, Queen's University, Kingston, ON K7L 3N6}

\begin{abstract}

Hubble Space Telescope imaging of 11 fields in M31 were reduced to
color-magnitude diagrams. The fields were chosen to sample all
galactocentric radii to 50 kpc ($\approx 9$ disk scale lengths, or
$>$99.9\% of the total light enclosed). The colors of the red giants
at each pointing map to an abundance distribution with errors of order
0.1 dex in [M/H]. The abundance distributions are all about the same
width, but show a mild gradient that flattens outside
$\sim$20 kpc. These distributions were weighted and summed with
the aid of a new surface brightness profile fit to obtain an abundance
distribution representative of the entirety of M31.  Since we expect
M31 to have retained most of its congenital gas and subsequently
accreted material, and since the present day gas mass fraction is
around 2\%, it must be a system near chemical maturity. This
``observed closed box'' is found to suffer from a lack of metal-poor
stars {\em and} metal-rich stars relative to the simplest closed-box
model in the same way as the solar neighborhood.  Models modified to
include inhomogeneous chemical enrichment, variable yield, or infall
all fit to within the uncertainties.  As noted elsewhere, stars in the outer
regions of M31 are a factor of ten more metal-rich than the Milky Way
halo, ten times more metal-rich than the dwarf spheroidals cospatial
with it, and more metal-rich than most of the globular clusters at the
same galactocentric radius. Difficulties of interpretation are greatly
eased if we posit that the M31 disk dominates over the halo at all
radii out to 50 kpc. In fact, scaling from current density models of
the Milky Way, one should not expect to see halo stars dominating over
disk stars until beyond our 50 kpc limit. A corollary conclusion is
that most published studies of the M31 ``halo'' are actually studies
of its disk.

\end{abstract}


\section{Introduction}

We have knowledge of the spiral Andromeda galaxy (M31; NGC 224) and
its elliptical companions M32 (NGC 221) and NGC 205 that rivals and in
some ways surpasses our knowledge of the Milky Way.  Comparing the two
galaxies can be a stringent test for theories of hierarchical galaxy
formation. The Andromeda Galaxy has about the same baryonic-plus-dark
mass as the Milky Way \citep{2000ApJ...540L...9E,2002MNRAS.337...34G}
but twice the number of globular clusters. Gas phase material
comprises 2\% of the baryonic total in M31, but about twice this in
the Milky Way \citep{vdb}. Further, the Milky Way is forming stars at
a greater rate, with a factor of $\sim$4 greater mass in ionised
hydrogen. Most of the gas in M31 is concentrated in a 10-kpc ``ring of
fire'' seen in molecular gas, atomic hydrogen, and dust, although OB
associations appear throughout the disk. \citet{vdk} finds that the
bulge of M31 contributes 25\% of the total light while Milky Way's
bulge contributes more like 12\%. (One should temper this result with
the realization that the Milky Way emits more light per unit mass than
does M31; in terms of mass the percentage contributions may be more
nearly equal. Also, our own $I$-band bulge-disk decomposition
discussed below yields a bulge contribution of 12\%, not 25\%.)

Globular cluster systems differ in the two galaxies despite both
having the same number of clusters per unit mass within a factor of
$\sim$2 \citep{vdb}. With a dividing line between metal poor and metal
rich placed at [M/H]~$=-1.0$, Milky Way metal rich globular clusters
are concentrated toward the center of the galaxy.  M31 has a larger
number of metal-rich globular clusters than the Milky Way. As in the
Milky Way, the metal rich M31 globular clusters lie preferentially
toward the center of the galaxy, but unlike the Milky Way, very
metal-rich globulars (of nearly solar abundance) are also found quite
far from the center. Unlike the Milky Way, a significant subset of M31
globular clusters appear to have strongly disk-like kinematics
\citep{morrison}.

Halo field stars in the Milky Way have an abundance distribution
similar to that of the globular clusters spatially coresident there:
with a peak at [M/H]~$\sim -1.5$ and a broad dispersion that matches
the simplest of chemical evolution models, the ``Simple model,'' that
assumes a closed box, full mixing, instantaneous recycling, and a
constant heavy element yield in every generation of stars. Halo field
stars in M31, if they do indeed belong to a halo population, are quite
different.  Among other authors, \citet{hh01,brown,rich,durrell,renda}, and \citet{grill} find from color-magnitude
diagrams that almost all of the field stars in the outer regions of
M31 are more metal-rich than 47 Tucanae (at [M/H]$\sim -0.8$), the
prototypical metal-rich disk globular cluster, with mean
abundances for the giants of [M/H]$\sim -0.5$.  This is almost
alarming: why should a galaxy of about the same mass have a halo
abundance a factor of ten higher? 

The aforementioned ``Simple Model'' of chemical evolution is a ``straw
man'' (easily knocked down) model that applies to a closed box of gas
that turns into stars. However, its simplicity, along with the
relative complexity of the alternative chemical evolution models,
makes it a compelling starting point. Furthermore, taken as a whole,
M31 inside a galactocentric radius of 50 kpc must be close to a closed
box in that it does not seem reasonable that it {\em permanently} lost
more than a pittance of its total gas mass to intergalactic
space. Models support near-total gas retention for high-mass galaxies
in non-cluster environments \citep{gov04}. Given that almost all of
its gas is either still in the M31 disk (2\% by mass) or has been
turned into stars (98\% by mass), then on a global level M31 is a
closed box or close enough to a true closed volume that outflow
should be negligible.

But M31 is a closed box only in the outflow direction. Infall in M31
could have occurred throughout its history, perhaps episodically where
each episode goes nearly to chemical completion before the next wave,
or perhaps steadily where the M31 gas fraction has been monotonically
decreasing to its minimum at the present epoch. In the latter case,
the Simple model might be expected to apply fairly well to the M31
closed box without modification. In the episodic case, one would have
a situation usually termed inhomogeneous chemical evolution. In
inhomogeneous chemical evolution, different patches of the galaxy mix
only slightly or not at all or are allowed to mix only between
enrichment episodes.

A final aspect of M31 that affects the discussion of its formation and
chemical evolution is the nature of its stellar halo. M31 does possess
a system of globular clusters that is spatially and kinematically
spheroidal, with a net rotation of only $80\pm 20$ km s$^{-1}$
\citep{huchra}. \citet{perrett} finds a net rotation of $138\pm 13$ km
s$^{-1}$ and velocity dispersion of $156\pm 6$ km s$^{-1}$ for a
sample of 321 clusters. \citet{racine} finds a globular cluster
surface density profile of $R^{-2}$ in the range $6 < R$(kpc) $< 22$,
but a steeper drop off, or even a cutoff, beyond this
range. \citet{vdb69} finds that the most metal-rich globular clusters
all have velocities of $\pm 100$ km s$^{-1}$ from the projected disk
velocity at their locations, indicating a more disk-like motion for
the metal-rich end of the cluster population. More extensive data
\citep{morrison} indicate that a significant fraction of M31 globular
clusters have thin-disk kinematics, are sprinkled over the entire
projected disk, and have abundances that span the entire metallicity
range. The M31 globular clusters have a spread of integrated colors
that covers about the same abundance range as Milky Way globular
clusters, but contains a more equal number of clusters in each
metallicity bin \citep{reed94}.

The surface brightness of field stars in the outer portions of M31 as
inferred from star counting follow, at least roughly, an
exponential $R^{1/4}$ profile \citep{vdb}. This is roughly $\rho (R)
\propto R^{-5}$ in the outer regions, as opposed to $R^{-3}$ as one
would infer for globular clusters \citep{racine}. 
But the continuity of the inner and outer parts of the minor
axis (illustrated, for example, in \citet{vdb}, figure 3.7) does not
demand a ``halo'' that is separate from an ``outer disk''. In fact
there is no evidence of a halo at all from the light profile of the
galaxy. Support for this viewpoint can be found from
the planetary nebulae kinematics of \citet{hurley}, who say, ``If M31
has a non-rotating, pressure-supported halo, we have yet to find it,
and it must be a very minor component of the galaxy.''

In following sections we discuss HST photometric observations that
sample galactocentric radii within 50 kpc, then we
present the inferred abundance distributions. (What we refer to as
abundance distributions are called by many ``MDFs,'' or metallicity
distribution functions, and are simply the fractions of stellar mass
observed over the range of abundance.) The observed distributions are
then weighted and summed to simulate a globally integrated ``closed
box'' abundance distribution for M31, which is then compared to solar
neighborhood observations and model predictions. We conclude with a
discussion of the constraints imposed upon the chemical evolution and
formation of the Andromeda galaxy and a discussion of the relative
dominance of disk vs. halo in the outer parts of the galaxy.

\section{Observations and Analysis}

A number of archived HST images were selected from the ``planned and
completed exposures'' list. The first selection pass was simply by
coordinates and whether the exposure was an image. The next step was
to look at whether exposure times were fairly long, and if images were
taken through two filters. For the outer fields it was clear that only
the WFPC2 had a sufficiently wide field of view to count enough stars
for meaningful results. The inner parts of M31 were observed by many
programs and we sampled from those, ensuring that we covered as wide a
range of radii as possible. Because inner fields suffer from crowding,
only the PC chip was used there. Outer fields suffer from sparse star
counts, so all available chips were utilised. Milky Way foreground
contamination as judged from the various color-magnitude diagrams was
negligible in all cases.

\begin{figure}  
\plotone{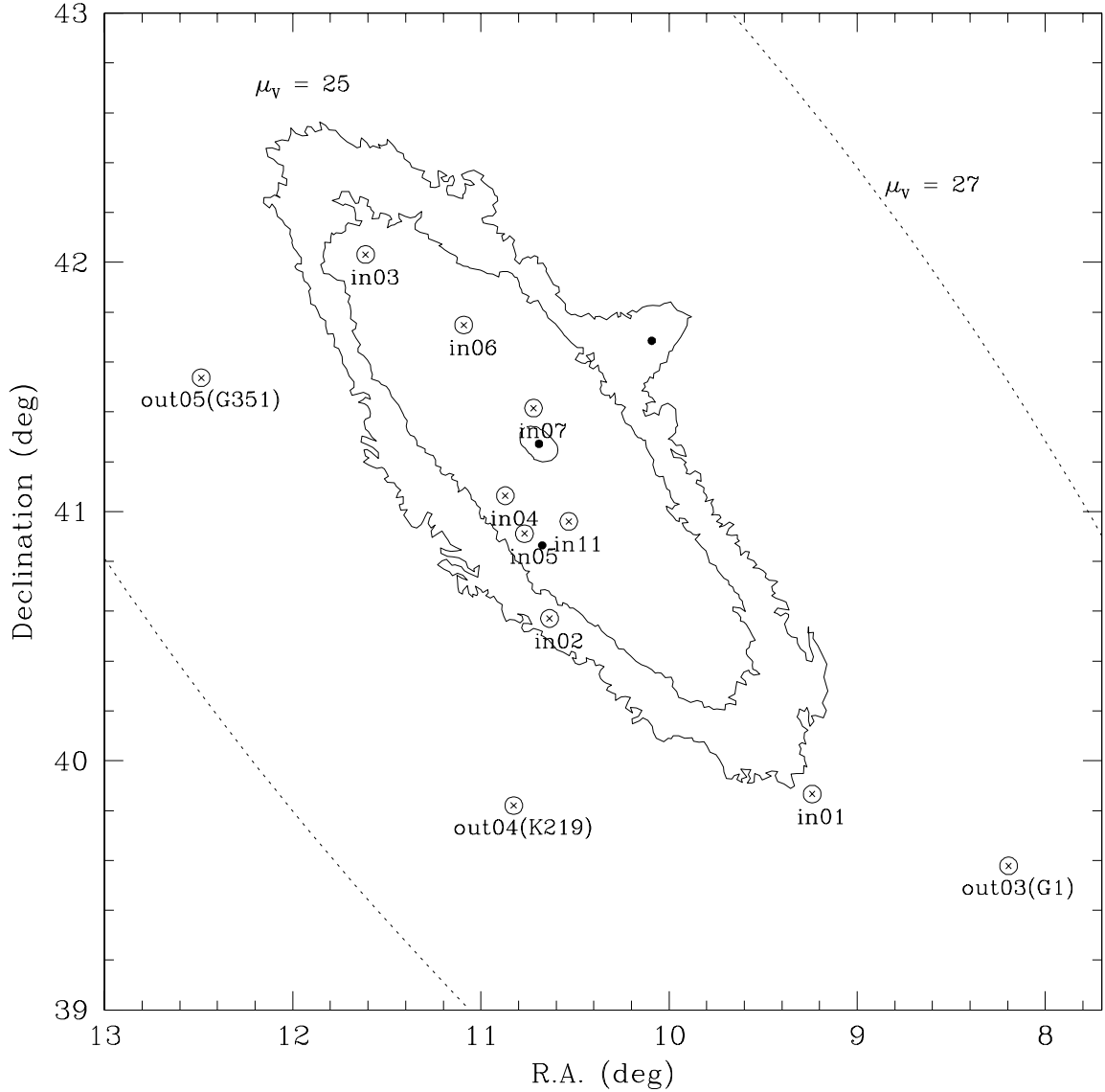}
\caption{ Program field locations, labeled by our internal ID, are
superimposed on M31 isophotes by \citet{hk}. The three ``out'' fields
happen to be globular cluster fields (with the cluster subtracted) and
the cluster names are in parentheses. Isophotes for $\mu_B = 25$,
$24$, and near-nuclear $21$ mag arcsec$^{-2}$ are shown. A rough
guess for $\mu_B = 27$ is sketched.  NGC 205 is marked by a dot
located inside the $\mu=25$ ``finger'' north of the M31 nucleus, and
M32 is marked by a dot just to the left of the ``in11'' label.
\label{figlocations}}
\end{figure}

The final fields are listed in Table \ref{tab1} and illustrated in
Figure \ref{figlocations}. The table lists (1) a project-internal
label for each field, (2) the WFPC2 filters used, (3) the semi-major
axis in arcsec from our radially-dependent elliptical isophotal
fitting, (4) the resultant galactocentric radius assuming an inclined
circular disk in kpc, (5) the galactocentric radius in the plane of
the sky (impact parameter) in kpc, (6) the original HST proposal
number, (7) the weighting factor used for combining the different
samples to make a composite M31 abundance distribution, (8) the number
of stars counted above the color-dependent cutoff magnitude, (9) the
abundance median, and (10) the abundance distribution FWHM. Many of
these quantities are discussed further below. The HST frames were
pipeline processed for basic flat fielding and photometric
calibration. Each data set consisted of some number of frames in two
filters to be coadded or ``drizzled'' to make final frames. Some of
the outer fields contained additional position shifts. In these cases,
each position was reduced separately. DAOPHOT \citep{stet} was used to
derive fluxes for as many stars as possible in each frame. Zero point
and aperture corrections were applied. The fluxes were corrected for
charge transfer efficiency effects via \citet{whitmore} and
transformed to Johnson-Cousins $V$ and $I$ via \citet{holtz}. A
reddening of E$(B-V)= 0.06$ mag \citep{schlegel} was assumed for all
fields.

Once color-magnitude diagrams are constructed, it is a straightforward
matter to overlay giant branches from isochrones of different
abundances. Stars that lie between isochrone giant branches are
assigned an abundance given by the mean of the flanking isochrones and
binned. A faint cutoff given by $M_I - 0.25(V-I) = -2.40$ was applied
so that photometric errors would not artificially broaden the inferred
abundance distribution too much. Finally, a correction for RGB stellar
lifetimes was applied to transform the observed star counts to mass
fractions (assuming an initial mass function for low-mass stars that
is constant as a function of heavy element abundance). More discussion
about the technique can be found in \citet{grill} and \citet{w04}.
The isochrones used were those of \citet{w94};
they have giant branches that fit $VI$ cluster observations better
than most existing sets over the entire range of abundance.  Despite the
faint cutoff, residual photometric error does slightly increase with
width of the resultant abundance distribution. We ignore this effect
because it is quite difficult to correctly implement such a
deconvolution given the geometry of the color-magnitude diagram and
also because rough estimates indicate that, with the observed widths
of roughly 0.6 dex for the final abundance distributions, the increase
of width due to photometric errors is only a few percent.

A possible problem worth recapping is the question of age and age
spreads. Age modulates giant star colors rather less than abundance in
a logarithmic sense. As \citet{w94} puts it, the line of null color
change is d[log(age)]/d[log(abundance)] $=-3/2$ so that (for example)
an age change from 8 Gyr to 10 Gyr can be exactly offset by an
abundance change of $-2/3 {\rm log} (10/8) = -0.06$ dex. If the
underlying age spread is 9 $\pm3$ Gyr the corresponding abundance
uncertainty is 0.08 dex. This error is slightly smaller than the one
we infer from random photometric uncertainties ($\sim$0.1 dex). The
best claim for intermediate-age subpopulations in the outer parts of
M31 is that of \citet{brown}, but ``intermediate age'' for those
authors is $\approx$6 Gyr. A subpopulation of this age is ``old'' for
our purposes; too old to cause any significant change in the abundance
distribution inferred from giant colors.

\section{Abundance Results}

\begin{figure}  
\plotone{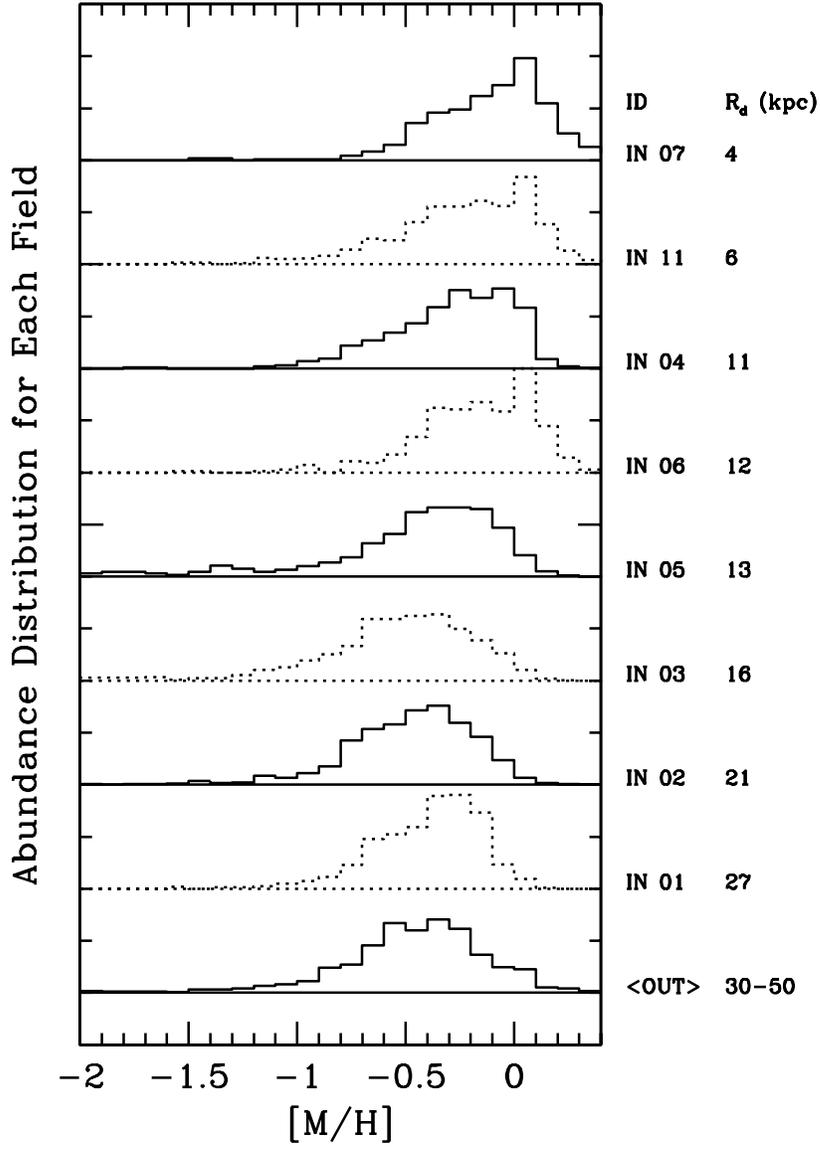}
\caption{ Derived abundance distributions for each field, except that
the outer fields have been averaged. Each distribution has been
normalised to unit integral, bins are 0.1
dex wide, and each vertical tick mark is 0.1. Labels for each field
appear to the right, along with the isophotal ellipse semi-major axis value
associated with each field. A mild abundance gradient is clearly seen.
\label{figstack}}
\end{figure}

Figure \ref{figstack} shows the resultant series of abundance
distributions for the various pointings, except that the star-starved
outer pointings have been summed. One can see a steady progression in
median abundance that goes from metal-rich in the inner disk to
relatively metal-poor in the outskirts. The widths of the
distributions do not change much from center to edge, and are similar
to the width of the abundance distribution in the solar neighborhood
(FWHM widths are listed in Table \ref{tab1}, derived by averaging the
top two bins to find the maximum, then linearly interpolating for the
half-widths).

\begin{figure}  
\plotone{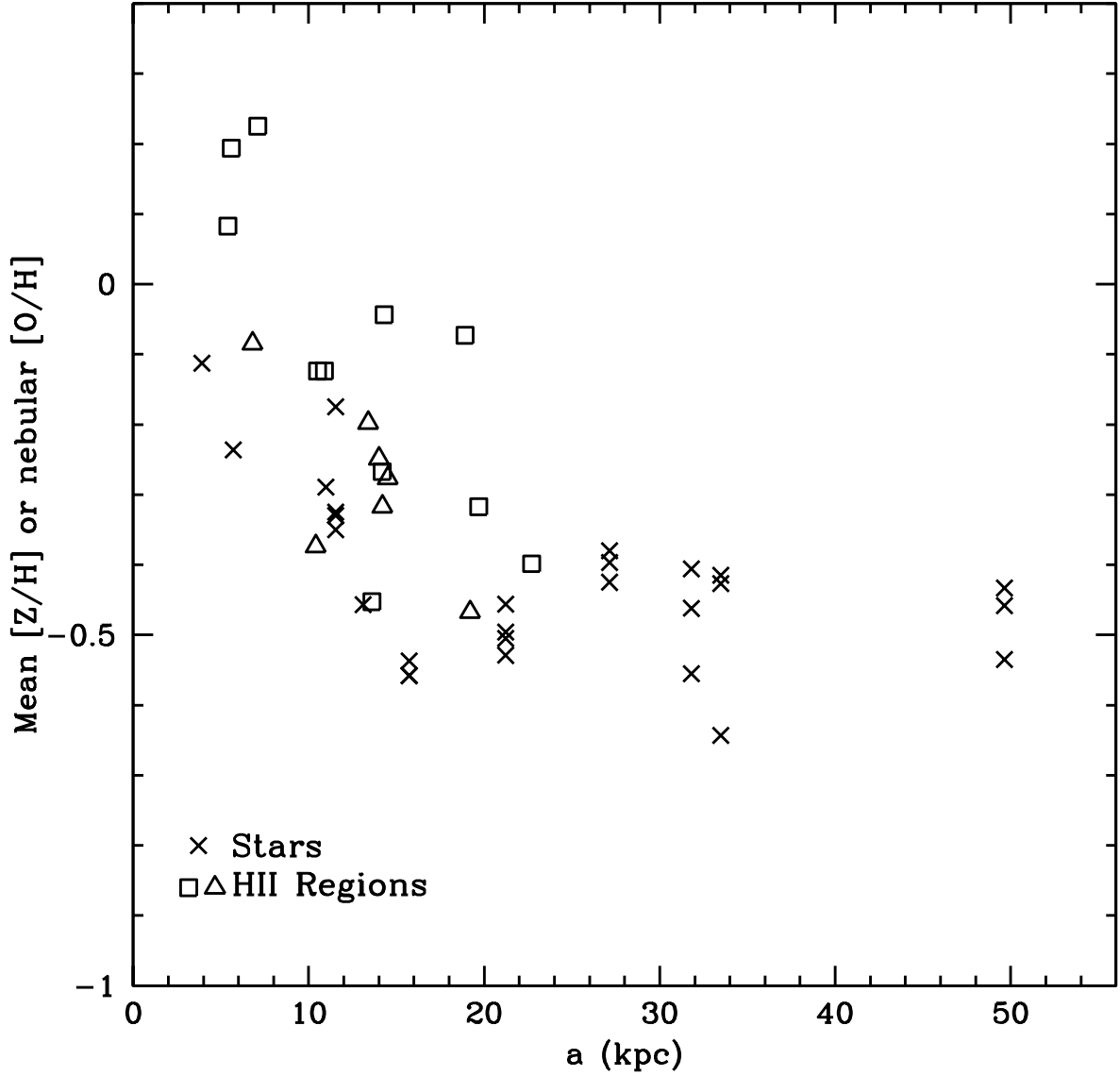}
\caption{ Median abundance for each field plotted as a function of 
galactocentric radius. Multiple points appearing at one radius 
indicate separate results for individual WFPC2 chips. Results from HII 
regions from \citet{blair} ({\em squares}) and \citet{denne} ({\em
triangles}) are also shown. 
\label{figgrad}}
\end{figure}

Figure \ref{figgrad} shows the median abundances for each field, but
split into median abundance for each chip (PC, WF1, WF2, or WF3) in
the cases where multiple chips were reduced. This is in order to gauge
the effects of (poor) counting statistics. The figure shows that the
outer disk-halo abundance holds relatively steady at [M/H]~$\approx
-0.5$. Gas-phase abundances from HII regions are also plotted on the
figure to show the present-day abundance. The nebular abundances at
least roughly agree with the slope of the gradient in the inner 25
kpc. To the extent to which one
can trust the nebular zero point, the gas is at an abundance roughly
0.2 dex higher than the median of the stars, which puts the gas on the
high-abundance end of the stellar abundance distributions. This is
what one would expect with Simple-model type chemical evolution with
good mixing. With inhomogeneous chemical evolution, one would expect
individual star formation regions to be distributed about the same way
as the fossil stars. We do not emphasise this point because the
nebular abundance zero point does have considerable uncertainty.

\begin{figure}  
\plotone{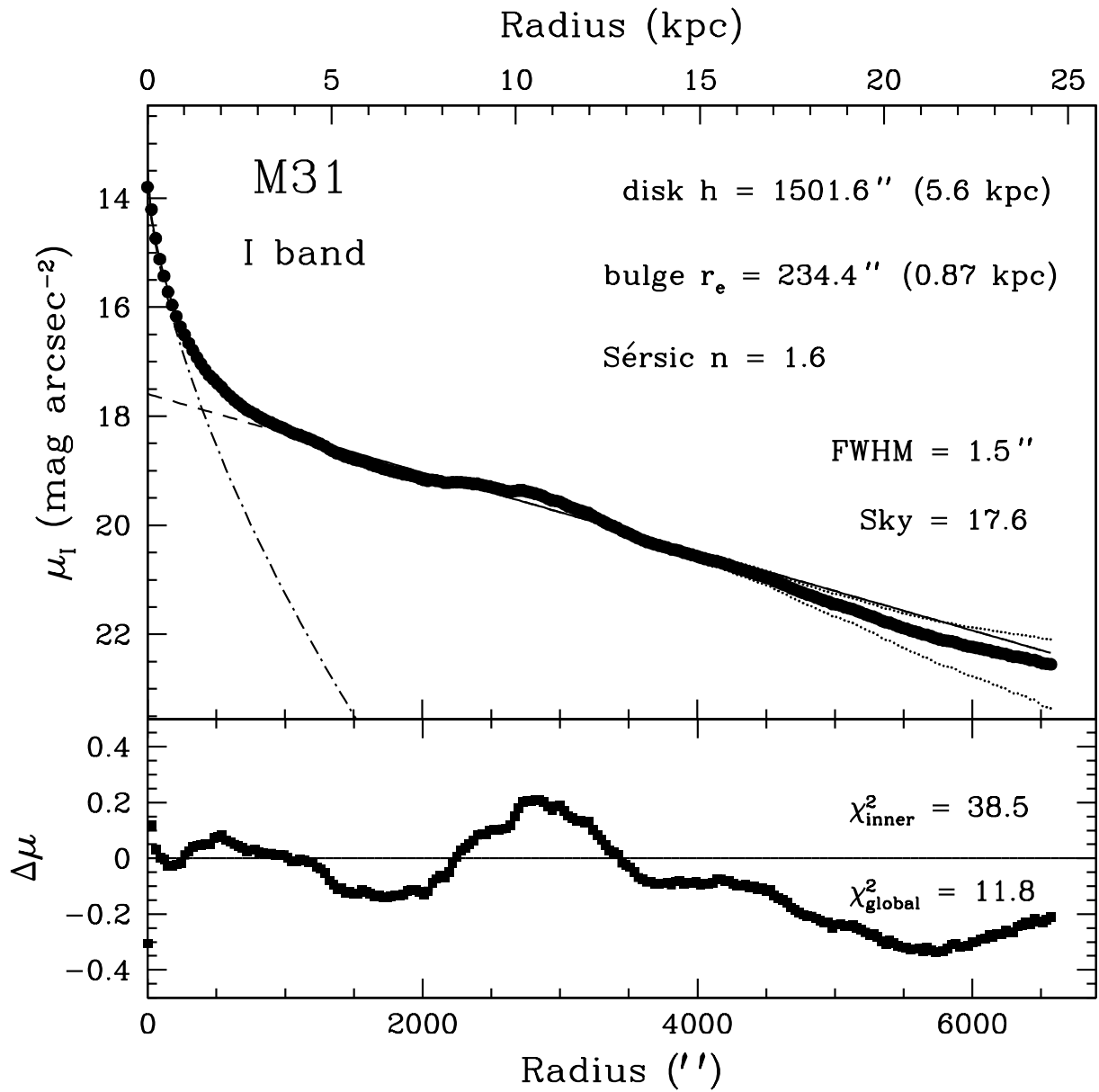}
\caption{Azimuthally-averaged surface brightness profile from elliptical 
isophotal fits to the \citet{choi02} $I$-band image.  
The bulge light was fit with a S\'ersic model of index $n=1.6$, 
effective radius, $r_e = 234\farcsec4$ (=0.87 kpc), and effective
$I$-band surface brightness $\mu_e = 16.6$ mag arcsec$^{-2}$.  The 
disk is described with an exponential profile of scale length
$h=1501\farcsec6$ (=5.6 kpc) and a central $I-$band surface brightness
$\mu_{0} = 17.6$ mag arcsec$^{-2}$.
The dotted lines show the surface brightness profile variations due to
a 1\% sky error.
Residuals from the fit are shown in the lower panel. 
The $\chi^2$ estimators and decomposition technique are
fully described in \citet{mac}.
\label{figfit}}
\end{figure}

The galactocentric radii of Figure \ref{figgrad} come from new
elliptical isophotal fits of an $I$-band mosaic \citep{choi02}. The
azimuthally-averaged surface brightness profile was extracted
following the method of \citet{1996ApJS..103..363C} and decomposed
into bulge and disk components. This is shown in Figure \ref{figfit}.
The disk brightness profile is very closely exponential except for a
stellar and gaseous ``ring'' beyond 10 kpc. The bulge brightness is
also closer to an exponential profile than a de Vaucouleurs, with a
S\'ersic index of $n=1.6$\footnote{In this notation, the exponential
and de Vaucouleurs profiles have a S\'ersic index equal to 1 and 4,
respectively.}.  The decomposition technique is fully described in
\citet{mac}. The surface brightness limit of the image was much too
bright to include regions that could be considered halo regions. The
sizes and orientations of the fitted ellipses were used to generate
the Table \ref{tab1} semi major axis column. Then, assuming an
inclined circular disk and a distance of 770 kpc \citep{freedman},
these semi major axis values were converted to a galactocentric
radius in kpc, also listed in Table \ref{tab1} and used for Figure
\ref{figgrad}. For the outer parts, we assumed an ellipticity of 0.679
and a position angle of $50^\circ$; the parameters of the last
reliable isophote.

\begin{figure}  
\plotone{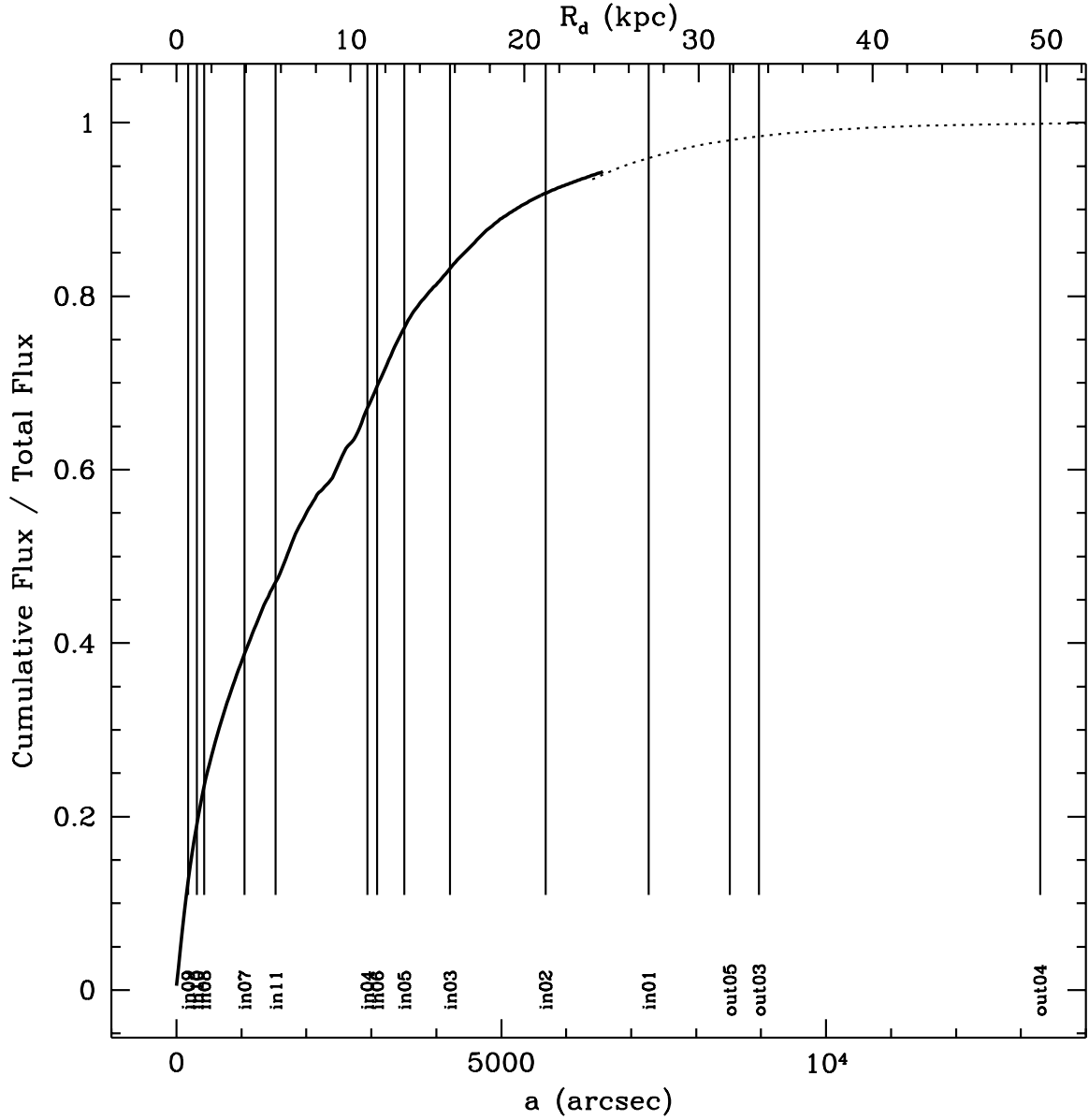}
\caption{ The measured cumulative surface brightness profile (inside
6400\arcsec; {\em solid line} ) given as a fraction of the total extrapolated flux and the assumed
extrapolation (outside 6400\arcsec; {\em dotted line}) are shown as a function of the
semi-major axes of fitted elliptical isophotes. Locations of the target
fields are also shown. Under the assumption of a circular disk
geometry, galactocentric radii are given on the top scale assuming a
distance of 770 kpc for M31. The three innermost fields were dropped
from the analysis due to severe stellar crowding that prevented
accurate photometry.
\label{figprofile}}
\end{figure}

The exponential isophotal fit was extrapolated to estimate the total
light of the galaxy. The observed cumulative light profile plus the
extrapolation is shown in Figure \ref{figprofile}. Locations of HST
fields are also shown in this figure, including the inner fields
(in08, in09, and in10) that were eventually dropped because stellar
crowding made deriving a reliable abundance histogram impossible. The
curve in Figure \ref{figprofile} was used to assign each pointing with
a weight representing the fraction of the galaxy ``covered'' by the
pointing. These weights are also listed in Table \ref{tab1}. One
notices that field ``In07'' is heavily weighted. This is because it
is the innermost field that was not beset with overcrowding
problems. This field lies outside the bulge. This means that the
bulge, as such, is not sampled. The fitted ratio of total bulge light
to total disk light is 0.137, so that even if the bulge abundance
distribution is very different from the rest of the galaxy it will not
affect the all-M31 abundance distribution much. Note that the
nuclear regions of M31 were studied in \citet{wdj}, who conclude that
M31 cannot have more than $\approx 5\%$ by mass contribution from
metal-poor stars, in harmony with our results where both field In07
and the all-M31 abundance distributions have 3\% of the mass in
populations with [M/H]$<-1$.

The weights for each field were then used to construct an all-M31 abundance
distribution from the individual sampled distributions. This global
abundance distribution should be a good approximation to the M31
closed box. The weighted-sum abundance distribution and each of the
individual distributions are tabulated in Table \ref{tab2} (except
that the outer fields are averaged). The distributions in Table \ref{tab2}
are normalised to a unit integral. The weighted sum
distribution has full-width-half-maximum (FWHM) of about 0.69 dex,
slightly broader than any of the individual abundance distributions. We
compare this global abundance distribution to local observation and
theory in the next section.

\section{Basic Chemical Evolution Models and Comparisons}

We can compare the global M31 abundance distribution to the solar
neighborhood abundance distribution and to simple models of chemical
evolution. We do not attempt to fit the full set of radially sampled
abundance distributions in this paper, but we recap the basics of
chemical evolution and try representative examples of all 
the classical schemes \citep{at76} for stepping away from the very simplest of
closed-box models, the Simple model.

\subsection{Analytic models}

The Simple model of chemical evolution assumes an isolated system, no
infall or outflow, of one zone, whose total mass is constant.  It
begins as pure gas with a metal abundance of $Z=0$ and is well
mixed at all times.  In addition, the IMF and nucleosynthetic yields
of the primary elements in the stars remain constant. 

To recap the Simple model [c.f. \citet{ss72,at76,bt}], we define
$M_h$ as the mass in gas phase made of heavy elements, $M_g$ the mass
of the gas, $M_s$ the mass in stars and define the abundance, or,
loosely, the metallicity, as $Z=M_h/M_g$.
If an incremental parcel of new stars is formed and only $\delta M_s$
remains locked in the stars, then the rest must be returned.  The
amount of gas returned by newly formed stars in heavy elements is $p\delta
M_s$, where $p$ is the ``nucleosynthetic yield'' in a given generation
of stars. One could assume that $p$ is a function of time or
metallicity, but the Simple model assumes a constant yield. We can now
calculate the change in the amount of metal-rich gas, which is the gas
returned minus the gas which formed the stars:
$ \delta M_h=p\ \delta M_s-Z\delta M_s =(p-Z)\delta M_s $.
Meanwhile, the change in the gas metallicity is
\begin{equation}
\delta Z=\delta \Bigl( {{M_h}\over{M_g}} \Bigr)
={{\delta M_h}\over{M_g}}- {{M_h}\over{M_g^2}} \delta M_g   
={{1}\over{M_g}} (\delta M_h-Z\delta M_g) .
\end{equation}
By combining equations and using the fact that mass is
conserved such that $\delta M_s$=$-\delta M_g$, we obtain
\begin{equation}
\delta Z={{1}\over{M_g}} [(p-Z)\delta M_s-Z\delta M_g]
={{1}\over{M_g}}[(-p+Z-Z)\delta M_g]
=-p\ {{\delta M_g}\over{M_g}} .
\end{equation}
If the yield $p$ is not a function of time, integration gives
\begin{equation}
Z(t)=-p\ {\rm ln}\Bigl[ {{M_g(t)}\over{M_g(0)}}\Bigr] .
\end{equation}
This expression for $Z$ is a function only of the yield and the gas
fraction, hence the term ``Simple'' for this scheme. Let us rework
this formula to give us what we observe: the number of stars at each
metallicity. If we let the gas
fraction approach chemical completion, how many stars of what
metallicity do we expect? The mass in stars of $Z$ less than $Z(t)$ at
some time $t$ is
\begin{equation}
M_s[<Z(t)]=M_g(0)-M_g(t)
            =M_g(0)\Bigr[ 1-{\rm exp}\Bigl( {{-Z(t)}\over{p}}\Bigr) \Bigr]
\end{equation}
so the differential fraction of mass at each $Z$ is
\begin{equation}
{{{\rm d}M_s}\over{{\rm d}Z}}={{{{\rm exp}(-Z/p)}}\over{p}} .
\end{equation}
But abundance is typically measured in terms of the logarithmic number
abundance like, for example, [Fe/H], or [M/H] if one lumps
all heavy elements together. We would like to recast equation (5) by
defining $F \approx $ [M/H] $ = {\rm log} (Z/Z_\odot)$.\footnote{Note
that this is not precisely correct. If ``heavy'' elements are those
other than hydrogen or helium, then one should account for the fact
that very metal-rich stars deplete their light elements and define $F=
{\rm log} {{Z/(1-Z)}\over{Z_\odot /(1 - Z_\odot)}}$. However, if one
considers stars from very metal-poor to a few times the solar
abundance of $Z_\sun \approx 0.02$, one can safely retain the
approximate form.} Using $Z=Z_\odot 10^F$ and d$Z= Z\ {\rm d}F/{\rm
log}(e)$, we arrive at 
\begin{equation}
dM={{Z_\sun 10^F}\over{p\ {\rm log}(e)}} {\rm exp}\Bigl[ -{{Z_\sun}\over{p}}10^F\Bigr] \ {\rm d}F .
\end{equation}
This distribution should be truncated at the gas fraction appropriate
for the system. In the case of M31, the present day gas fraction is
${{M_g(t)} \over {M_g(0)}} = 0.02$. 

\begin{figure}  
\plotone{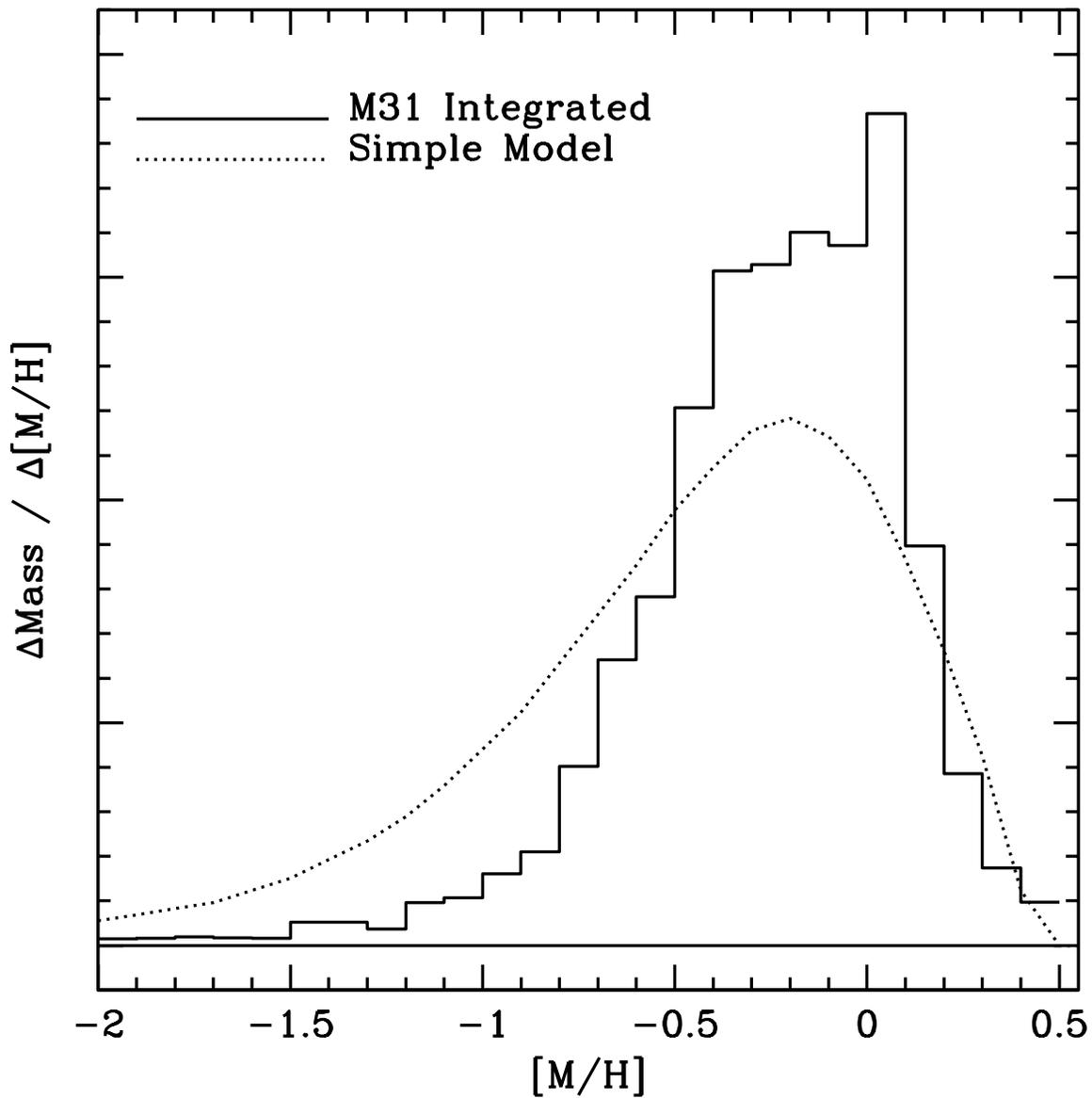}
\caption{The global, weighted-sum M31 abundance distribution is
compared to the Simple model with yield [M/H]$_p \approx $log[$p/Z_\sun] =  -0.2$ and a gas
fraction of 2\%. 
\label{fighist1}}
\end{figure}

In Figure
\ref{fighist1} the Simple model and the global-M31 abundance
distribution are compared. Note that M31's small present-day gas
fraction means that chemical evolution in the Simple model is almost
complete, and only a small modification in the most metal-rich bin is
needed to account for the truncation of the model distribution. What
is evident immediately is that the Simple model abundance distribution
is wider, with an extended tail toward metal-poor stars. This is the
same mismatch as the ``G dwarf problem'' seen in the solar
neighborhood. 

\begin{figure}  
\plotone{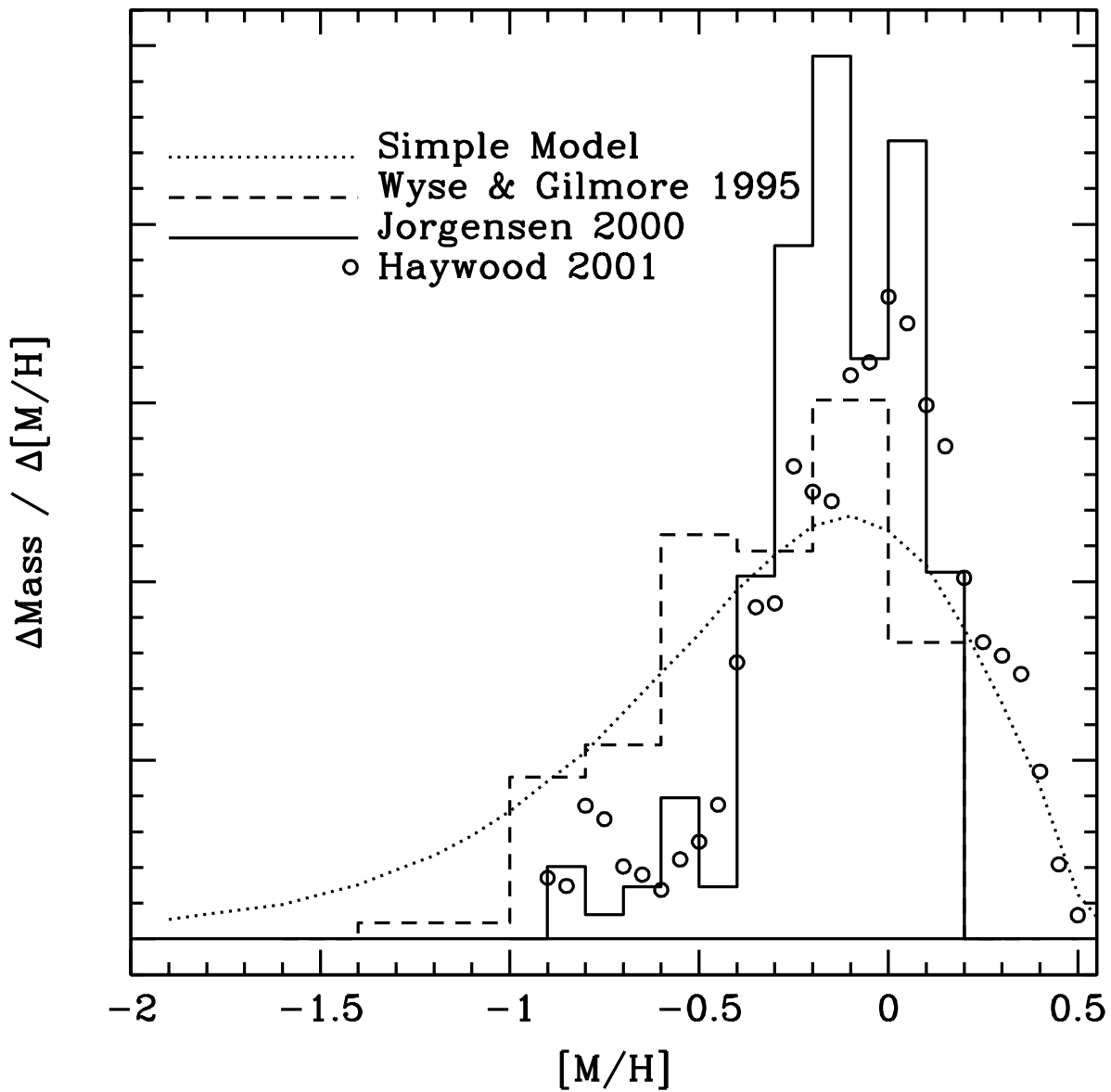}
\caption{A Simple model with yield [M/H]$_p = -0.1$ and gas fraction of
2\% is compared to three independent literature abundance
distributions derived for the solar cylinder. The literature
references are \citet{wyse}, \citet{jorgensen}, and \citet{hay01}.
\label{fighist2}}
\end{figure}

In Figure \ref{fighist2} a Simple model (with a yield 0.1 dex higher
than Figure \ref{fighist1})
is compared to various solar neighborhood data sets. The broadest of
these is \citet{wyse} (FWHM 0.75 dex), and the narrowest is
\citet{jorgensen} (FWHM 0.47 dex). These widths more than bracket our
derived M31 abundance distribution widths listed in Table \ref{tab1}
and the closed-box M31, which has a FWHM of 0.69 dex. The
\citet{hay01} distribution is intermediate in width (FWHM 0.61 dex),
but is shifted toward metal-rich stars. \citet{wyse} 
emphasised corrections to augment the metal-poor end of the
distribution to account for thick disk and halo stars while 
\citet{hay01} emphasised selection effects against metal-rich
stars. The differences between authors are large, illustrating that
the solar neighborhood data are subject to a fair amount of
interpretation. All of the distributions, however, are more narrow
than the Simple model (FWHM 0.87 dex), so the solar neighborhood has a
G dwarf problem for all data sets.

Any number of modifications to the Simple model can make it fit the
data better.

The first scheme is to have a nucleosynthetic yield that starts
high, then decreases with increasing abundance (an obvious way to make
chemical evolution proceed more quickly in metal-poor
regimes). Whether or not this applies in the real universe is
unknown, but many authors have been driven to consider the idea of a
variable initial mass function that might drive a variable yield 
\citep{bresolin, cena, cenb, scanna, schneider, chabrier, larson,
padoan97, chia00, chiosi98, weid03}. Alternatively, supernovae
themselves may have a different character at lower abundance.

We modify the Simple
model to have a yield that begins high and then decreases with increasing
$Z$ as $p = {{p_0}\over {Z + \epsilon}} $. We have not seen this particular
variant in the literature; we call it the ``rational decreasing yield
model'', but really it is not a physical model and serves only to
illustrate how easily one can fit the observations. This scheme is a
three-parameter model with the gas fraction remaining as the primary
parameter, but the yield now parameterised with $p_0$ and
$\epsilon$. Substituted into equation (2) and integrated, the
expression for $p$ yields a quadratic with positive root
\begin{equation}
Z(t) = -\epsilon + (\epsilon^2 - 2 p_0 {\rm ln}\Bigl[
{{M_g(t)}\over{M_g(0)}} \Bigr] )^{1/2} .
\end{equation}
The analog of equation (4) for this model is
\begin{equation}
M_s[<Z(t)] = M_g(0)\Biggl( 1 - {\rm exp}\Bigl[ - {{(Z+\epsilon)^2 -\epsilon^2 }\over{2p_0}} \Bigr] \Biggr) .
\end{equation}
The analog of equation (5) is
\begin{equation}
{{{\rm d}M_s}\over{{\rm d}Z}} = {{Z+\epsilon}\over{p_0}} {\rm exp}\Bigl[- {{(Z+\epsilon)^2 + \epsilon^2}\over{2p_0}} \Bigr] .
\end{equation}
And converting to ten-based logarithmic units yields
\begin{equation}
{\rm d}M_s = {{Z_\sun 10^F + \epsilon}\over{p_0}}\ {\rm exp}\Bigl[ -
{{(Z_\sun10^F + \epsilon)^2 - \epsilon^2}\over{2p_0}} \Bigr] {{Z_\sun
10^F}\over{{\rm log}(e)}} {\rm d}F .
\end{equation}

\begin{figure}  
\plotone{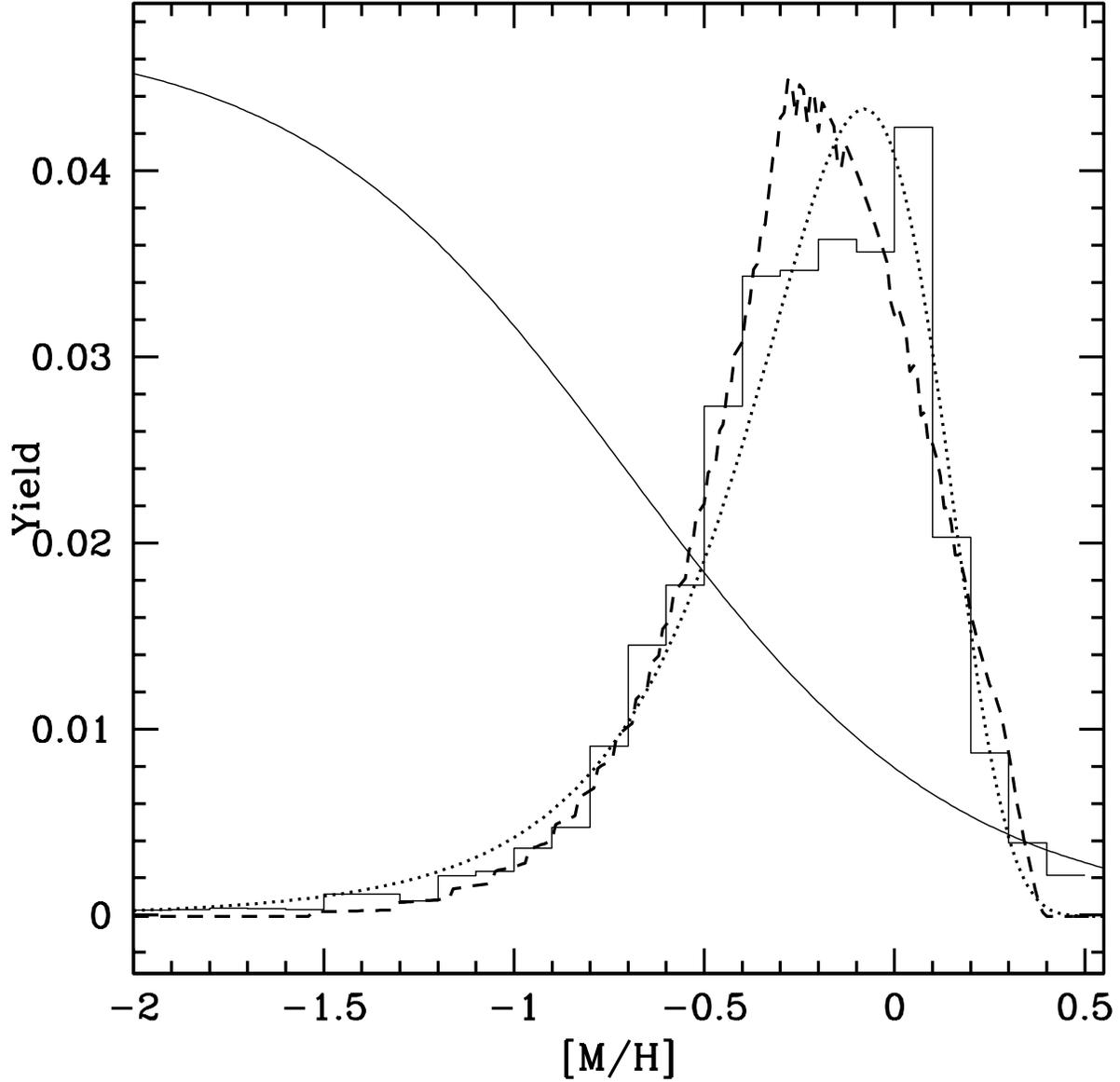}
\caption{The M31 abundance distribution is compared to the rational
decreasing yield model with $p_0 = 0.00019$ and $\epsilon = 0.004$
({\em dotted line}). The thin line and the y-axis scale illustrate the yield
$p$ computed with these assumptions. The slightly ragged dashed line
is a \citet{malinie93} model with parameters as in their paper except
that the initial metallicity is zero and the $f$ parameter has been
set to 0.075 so that the final gas fraction is 0.02.
\label{fighist4}}
\end{figure}

Figure \ref{fighist4} illustrates the improvement in fit that
can be achieved with a variable yield. The parameter $p_0$ controls
the abundance of the peak of the distribution, while $\epsilon$
controls the width. The scale on the $y$ axis and the thin decreasing
line represent the variable yield, computed with $p_0 = 0.00019$ and
$\epsilon = 0.004$. The shape of the resultant abundance distribution
is not sensitive to modest changes in the parameters; a good feature
of this modified model. Another good feature is the plateau at the
very metal-poor end, but there are mild mismatches at various points
as well. The width is FWHM = 0.62 dex for the parameters listed.

\subsection{Inhomogeneous Enrichment Models}

Suppose that, in a cube of gas, enrichment occurs in patches scattered
throughout the cube. These patches may or may not overlap spatially,
so that if little mixing occurs then a variety of abundances will
coexist in the gas. Such inhomogeneous chemical
evolution schemes been studied in several works [e.g. \citet{tin75},
\citet{searle77}, and \citet{malinie93} ]. 

First, we compare our M31
composite abundance distribution to a model with zero mixing
\citep{oey00}. This model has five parameters: the volume filling
fraction $Q$ of the enriched bubbles, the number $n$ of enrichment
generations, the mean $a$ and standard deviation $\sigma$ of the
abundance distribution typical of a collection of local star formation
events, and the present day gas fraction $\mu = M_g(t)/M_g(0)$. 
We assume that each generation of star formation is
described by a Gaussian abundance distribution; i.e., we stop at
Oey's equation (8) rather than going on through equation
(13).

\begin{figure}  
\plotone{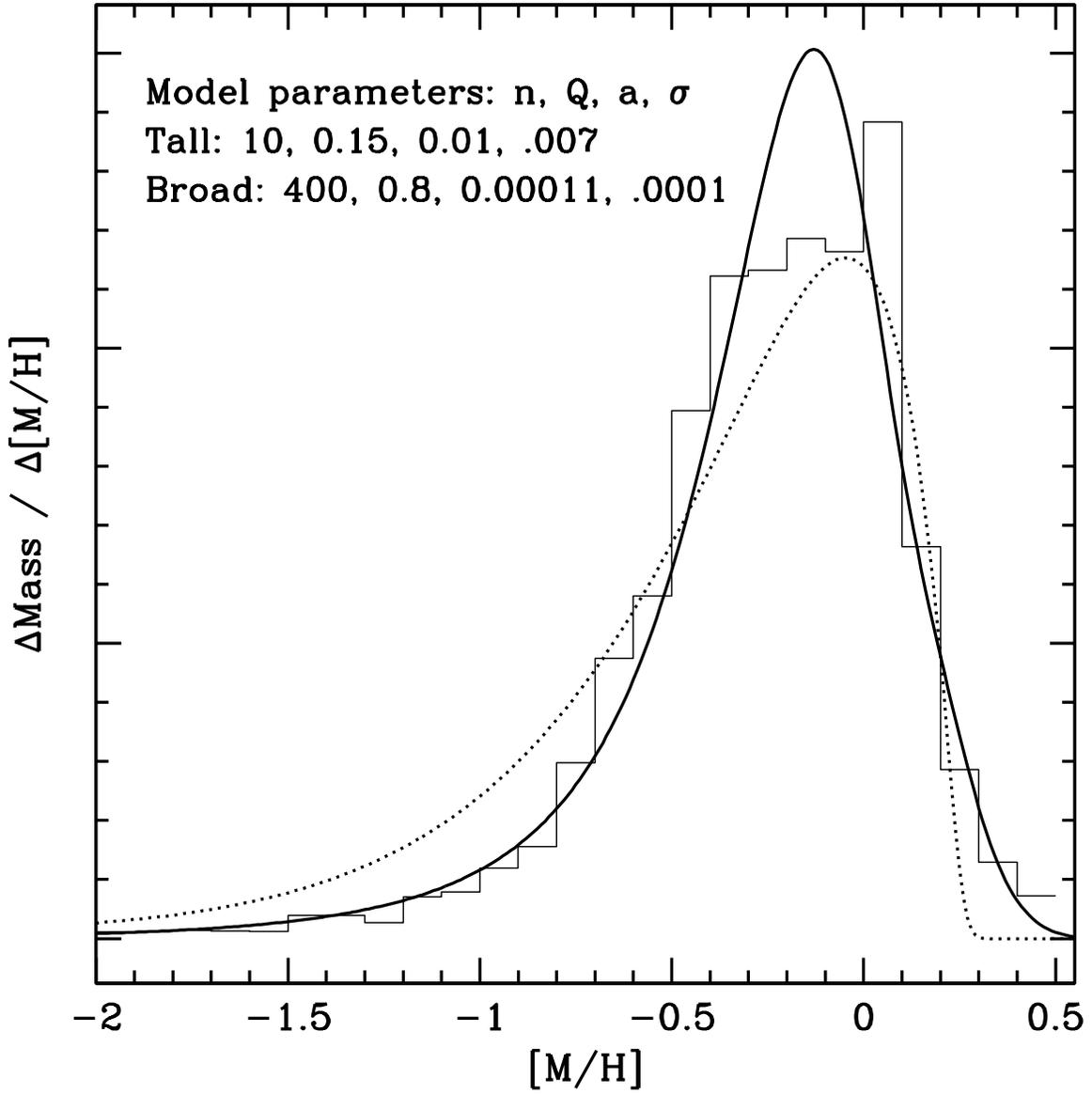}
\caption{The M31 abundance distribution is compared to two 
inhomogeneous models computed using the formalism of \citet{oey00} 
with $\mu = 0.02$ and other parameters as noted in the plot.
\label{fighist3}}
\end{figure}

Some \citet{oey00} models are plotted in Figure \ref{fighist3}, with
parameters listed in the figure itself. The broad model (FWHM 0.76
dex) with $n=$ 400 and $Q=0.8$ is similar to what is plotted in her
Figure 4 that is meant to match the solar neighborhood data, although
Oey uses a different gas fraction.  The other, narrower distribution
(FWHM 0.55 dex) has the number of events multiplied by the volume
filling factor of order one ($nQ\sim 1$).  We worry that this
corner of parameter space is not ideally suited for spiral galaxies with
long star formation histories, but the model does fit quite nicely.

We also compare the M31 abundance distribution to the inhomogeneous
chemical evolution model of \citet{malinie93} in which mixing is
allowed between, but not during, inhomogeneous star formation events.
This model was built to emulate the solar neighborhood situation and,
like the Oey model, has 5 parameters; the number of mixing events $N$,
the fraction of gas consumed in each event $f$, an initial metallicity
$Z_0$, a metallicity dispersion $\delta Z$, and a yield $y$. Figure
\ref{fighist4} illustrates the match of the \citet{malinie93} model
(dashed line) if the initial metallicity $Z_0=0$ and the fraction of
gas consumed in each event ($f$) is set so that the final gas fraction
is 0.02 instead of 0.2 as in their paper. Other parameters are left as
in \citet{malinie93} ($N=100$, $\delta Z=0.01$, and $y=0.72
Z_\sun$). Setting the number of events $N$ equal to 100 implies that
there was quite a lot of mixing in the history of M31. The Malinie
model (FWHM 0.63 dex) is quite successful at roughly matching the M31
closed-box abundance distribution (FWHM 0.69). The only parameter
changes from the Milky Way case were the expected ones: an adjustment
for a lower gas fraction in M31 and an initial metallicity of zero
that is appropriate for a closed box situation.

\subsection{Infall}

We can also massage the shape of the abundance distribution by
feeding gas into the box from outside, even as chemical evolution
proceeds within the box. This could be done with complete mixing or
with inhomogeneity. We illustrate the former option using the
parameterization of \citet{hh02}. This is a five-parameter model: the
metallicity of the infalling gas, a constant yield, and a constant gas
consumption rate, expressed as the fraction of gas consumed per time
step, $e$. The rate of infall is constant for some time $\tau_1$, then
decays exponentially as exp$(-t/\tau_2)$, where $t$ is an integral
number of time steps. 

\begin{figure}  
\plotone{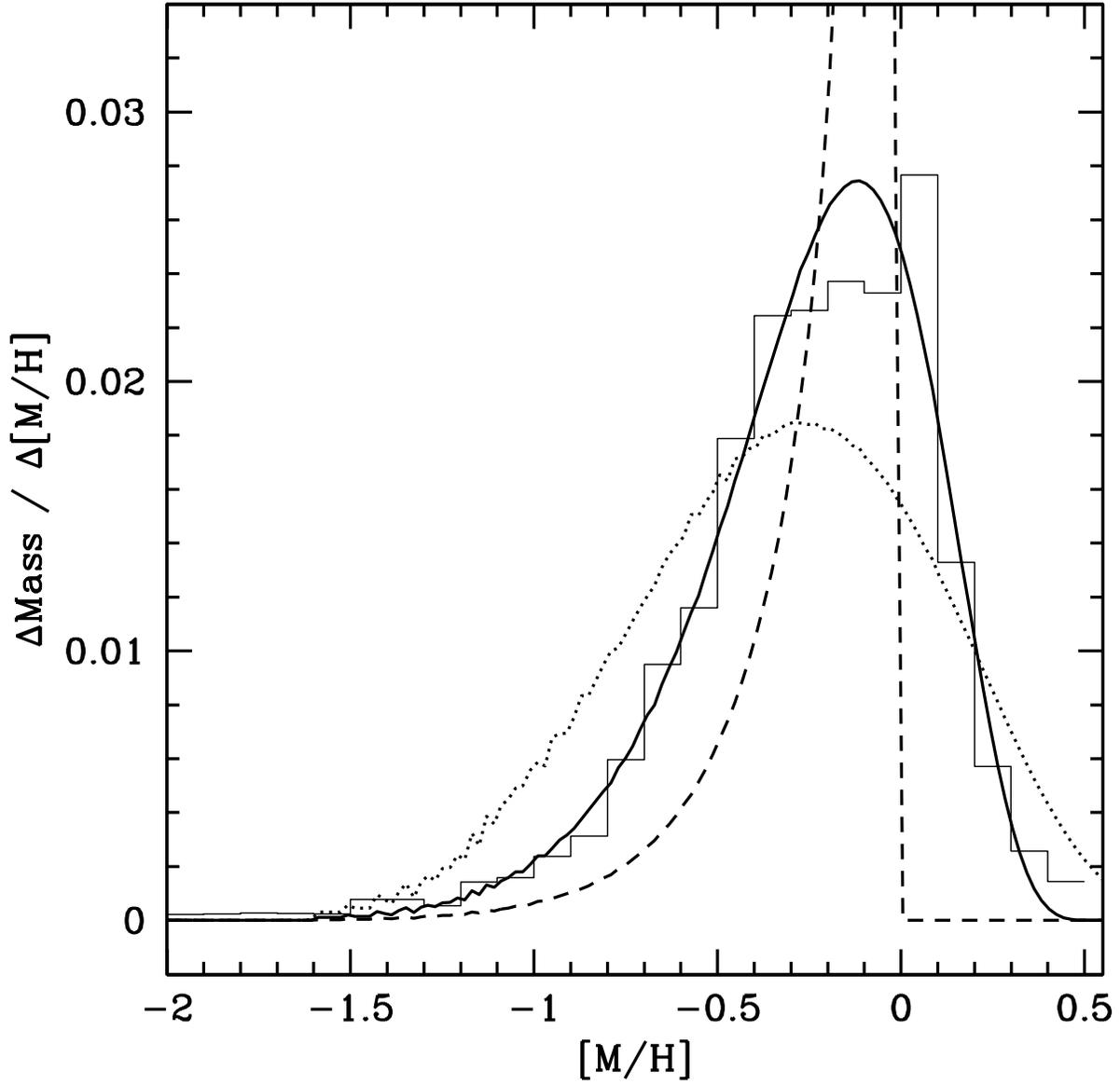}
\caption{The M31 abundance distribution ({\em thin line}) compared to
infall models computed using the parameterization of \citet{hh02} with
yield = 0.7$Z_\sun$, fraction of gas consumed per time step $e=0.02$,
and $\tau_1 = 10$ time steps of constant infall before exponential
decline sets in. The subsequent decay time constant is $\tau_2=65$
time steps for the model that matches the observation ({\em bold}),
and $\tau_2$ is set to four times longer for the narrow model whose
peak is off the graph (at 0.14; {\em dashed}), and four times shorter
for the broad model ({\em dotted}).
\label{fighist5}}
\end{figure}

Figure \ref{fighist5} shows representative models compared to the M31
data. The most sensitive input parameters are the yield and the
$\tau_2$ infall timescale. In Figure \ref{fighist5} a model that fits
the data (with FWHM=0.68) is plotted along with models in which the
decay timescale is varied by a factor of 1/4 and 4.  A more constant
star formation rate approaches a delta function shape in the abundance
distribution, while a very short star formation time makes the
distribution very broad. While the yield might be a universal
constant, infall timescales are not, so this model gives a prediction
that galaxies with short formation times or that formed in bursts
(like at least some elliptical galaxies) should have much broader
abundance distributions than galaxies with long formation timescales
(Sc or Sd or LSB galaxies) {\em unless} infall rate is directly
proportional to star formation rate; the model abundance distribution
is invariant in this case. Due to the sensitivity of the model to the
infall timescale, the infall rate would have to be proportional to the
star formation rate to a high degree of accuracy across Hubble types
and across quiescent or bursty star formation modes in order for it to
apply to real galaxies, which tend to have very similar abundance
distributions: \citet{wdj} note that, so far, every galaxy studied
that is bigger than M32 seems to suffer a ``G dwarf problem'' of
having a narrow abundance distribution compared to the Simple model.
[Objects include the Milky Way, M32, M31, and the nuclear regions of
elliptical galaxies.  The outer regions of disturbed giant elliptical
NGC 5128 have also been shown to have a narrow abundance distribution
with a FWHM $\sim0.65$ dex \citep{hh02,rej04,rej05}]. This argues for a
universal phenomenon rather than a special case scenario.  Observed
star formation rates in gas-rich galaxies range from $< 10^{-2}$
M$_\sun$ yr$^{-1}$ in Sc/Sd/LSB spirals \citep{burk01} to $> 10^3$
M$_\sun$ yr$^{-1}$ in ultraluminous infrared galaxies
\citep{smail}. One would have to require that infall rate and star
formation rate are synchronized to about a factor of two over this
range in order for this exact infall model to apply. On the other
hand, infall {\em must} be an important component of galactic
evolution.  Plausibly, alternative methods of parameterization
may show less volatility.

\section{Discussion}

\subsection{Closed Box Models}

In summary, the comparisons of the observed M31 closed box abundance
distribution with chemical evolution models showed that, while the
Simple model is too broad, all kinds of slightly modified chemical
evolution schemes fit the narrower observed abundance distribution of
closed-box M31 stars.
Combined with the result of \citet{wdj} that elliptical
(i.e. spheroid-dominated) galaxies also have a narrow abundance
distribution compared to the Simple model, the ``narrowness'' of the
abundance distribution appears to be a universal phenomenon in all
large galaxies, not just for disk galaxies. This is circumstantial
evidence that the abundance distribution is generically narrower than
the Simple model suggests. 

Since outflow is ruled out for closed boxes, the remaining classical
solutions to this problem \citep{at76} are (1) prompt initial
enrichment, (2) infall without outflow, (3) dropping the assumption of
instantaneous recycling, i.e. allowing for time-delayed yields and
allowing inhomogeneous enrichment. Our $Z$-dependent yield is an
implementation of (1), and it works fairly well except that the
parameterisation is not physical.  Scheme (2), infall, easily matches
almost any abundance distribution with suitable fine-tuning of yield,
mixing, and infall parameters, as we show with the full-mixing
\citet{hh02} parameterization.  The problem with infall models as
applied to galactic closed boxes is that the output abundance
distribution is very sensitive to the infall timescale, so that the
infall timescale would have to be nearly the same for ellipticals as
it is for spirals since they have very similar abundance
distributions. This objection is nullified if the infall timescale is
always directly proportional to the star formation rate to about a
factor of two in all star formation environments.  Scheme (3) was
partially explored with inhomogeneous models. The easiest match
included substantial mixing of gas during chemical evolution.  Delayed
injection of heavy elements serves to add metal-poor stars to the
system, and so would make the fit worse if it were included. We are
therefore left with a theme often voiced about chemical evolution;
that there is a uniqueness problem. Several models with different
assumptions can match the data quite well \citep{tosi}.

\subsection{Disk Dominance}

As a prelude to further discussion, we ask ``at what radius do we expect M31
halo stars to dominate over disk stars?'' assuming that the M31 halo
resembles that of the Milky Way. Disks are usually modelled with an
exponential profile of the form $I=I_0\ {\rm exp}(-R/h)$, where $h$ is the
scale length and $I$ is a surface brightness, not a density. The Milky
Way halo is found to follow a power law density $\rho = \rho_0
R^{-3.5}$ outside of 5 kpc \citep{vdb}. Provided we can trust these
functional forms, an exponential will fade faster than a power law, so
a halo will always dominate outside some radius.

\begin{figure}  
\plotone{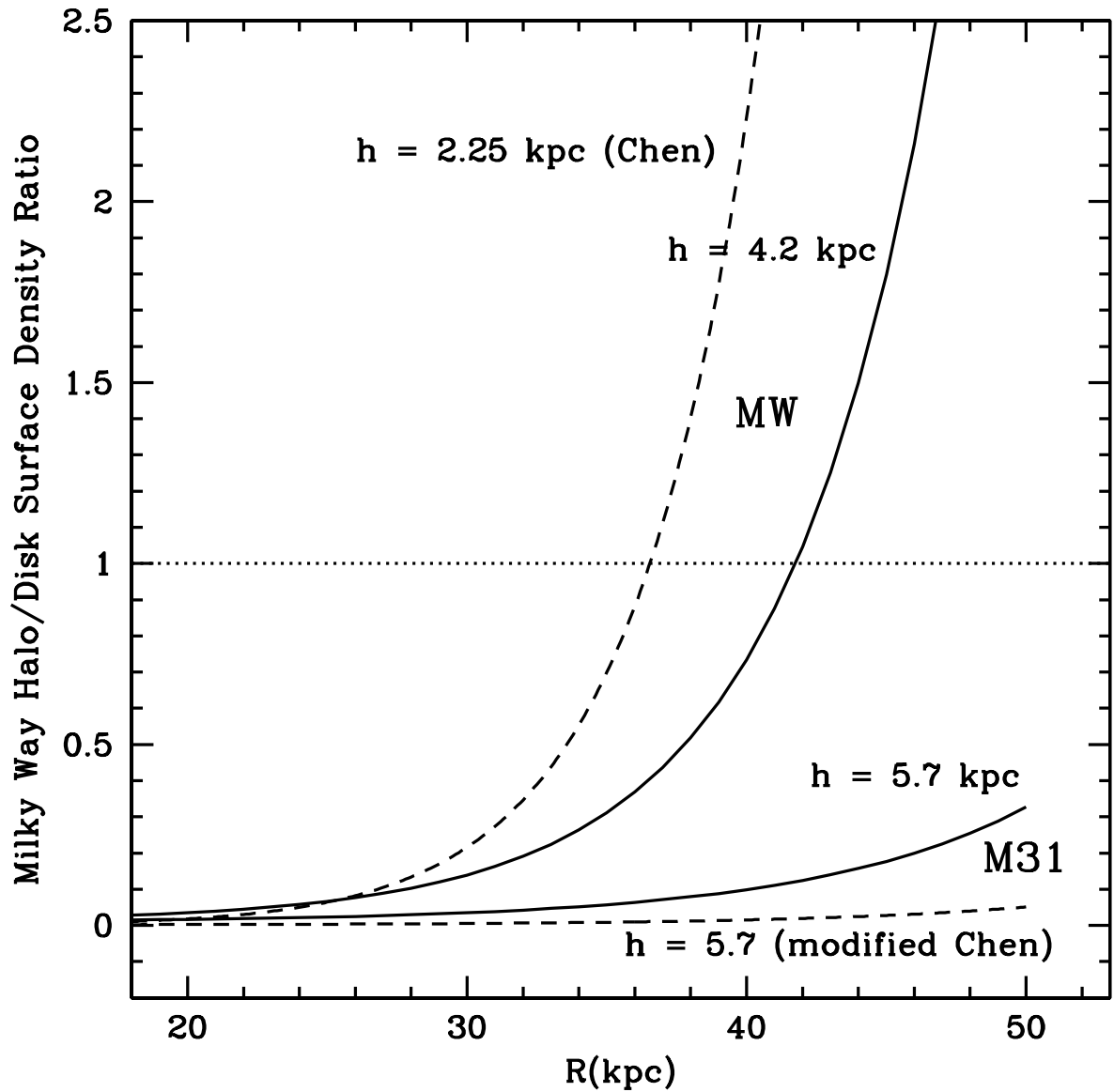}
\caption{ The ratio of halo to disk surface density (that is,
projected on the $x$, $y$ plane) for some models of the Milky
Way. Values greater than one indicate that most stars seen by an
observer distant from us and looking down on the disk would be halo
stars, and values less than one indicate that disk stars dominate the
light. The \citet{chen} model has a scale length of 2.25 kpc, and a
model based on \citet{sandage} plus \citet{vdb}'s preferred scale
length of 4.2 kpc is also shown. The Milky Way in these models has a
halo that becomes visible somewhere between 35 and 45 kpc. Both of
these models are modified to emulate M31 by increasing the disk scale
length to 5.7 kpc and (conservatively) doubling the density of the halo. The
figure shows that, with a Milky Way-style halo, we should expect to
see very few halo stars at 50 kpc from the center of M31.
\label{figratio}}
\end{figure}

To evaluate the surface density of the halo, normalised to the local
surface density of the disk, we use the results of \citet{sandage},
integrating his Table 1 (kinematically derived densities as a function
of height above the disk) and adopting local density ratios for
disk:thick disk:halo of 500:30:1 (and for our purposes we simply
subsume the thick disk into the disk). The problem is then fully
specified except for the disk scale length $h$. Figure \ref{figratio}
shows the prediction for the $V$-band Milky Way scale length $h=4.2$
kpc \citep{vdb}. Van den Bergh notes that estimates of 2 to 6 kpc for
the Milky Way scale length $h$ exist in the literature, part of this
spread being a function of the wavelength studied.  The \citet{chen}
models are also shown in Figure \ref{figratio} exactly as integrated
from the published density model (including a shallower halo density
power law of $-2.5$ and disk scale length of $h=2.25$ kpc). These
results indicate that, in the Milky Way, halo stars become more common
than disk stars at galactocentric radii of 35 kpc $< R <$ 45 kpc.

The scale length of M31 is more secure than that of the Milky
Way. \citet{vdb} quotes $h=5.7$ kpc for the $V$-band, and our own
$I$-band decomposition gives $h=5.6$ kpc [or $5.7$ kpc if the slightly
longer distance of 784 kpc \citep{stan} is used]. For a scale length
of this size, and also doubling the density of the M31 halo because
M31 has twice the number of globular clusters, Figure \ref{figratio}
indicates that we should {\em not} expect to see the halo dominate at
any radius that we sample, and that halo stars should become
relatively numerous only outside a radius of 50 kpc from the center of
M31.

This should be slightly tempered by a geometrical consideration: If
the halo is spherical, a line of sight intersects the halo at an
impact parameter $R_h \approx D\theta$, where $D$ is distance and
$\theta$ is the angle between the line of sight and the M31
nucleus. The impact parameter is the same as the galactocentric radius
in the inclined disk ($R_d$) only along the major axis of the
isophote. For most lines of sight, the projected disk is foreshortened
so that $R_h < R_d$, and the foreshortening is at a maximum along the
minor axis of an isophote \citep{wk87}.  For gauging this possible
amplification effect, we list both radii in Table \ref{tab1} [the
amplification of the halo is roughly $(R_d/R_h)^{2.5}$ assuming the
halo density profile is $\rho \propto R^{3.5}$].  We find that fields
near the minor axis can have the halo boosted by a factor of order ten
from this effect. Of course, if the halo is flattened instead of
spherical (as the Milky Way's is believed to be) the boost rapidly
diminishes. And there is evidence from globular clusters that the halo
density falls off faster than $\rho \propto R^{-3.5}$ outside of 30
kpc \citep{racine}. It is possible that, with the exception of nearly
edge-on spirals where projection effects dominate, the disk stars
outnumber the halo stars at all radii. We note that the metallicity
remains about [M/H] $\sim -0.5$ even in fields as far as $R_h = 30$
kpc \citep{dur04,durrell}.

If we are seeing disk stars in all fields, then explaining the relatively
metal-rich mean abundance of [M/H]$\approx -0.5$ becomes quite natural
compared to trying to explain why the halo of M31 is a factor
of ten more metal-rich than the Milky Way halo. If we are seeing disk
stars, then the M31 halo can be presumed to be alive and well and very
similar to the Milky Way halo, which, after all, is supposed to
compose less than 2\% of the mass of the Galaxy. Alive and well, but,
as yet, unseen except for the metal-poor-selected stars of \citet{rg02}.

Many authors have called the outer regions of M31 ``halo,'' but for
the most part this is a term of convenience, and not indicative of
true pressure-supported halo status since we are almost completely
ignorant of the kinematics of these stars as of this
writing. Parantheticallly, our ignorance is about to be relieved as
large telescopes are beginning to measure many velocities of giant
stars in the uncrowded outer regions of M31 [cf. \citet{rg02}]. It is
interesting to wonder how interpretations change if disk status is
assumed. \citet{brown} find an ``intermediate age'' subpopulation in a
field at 11 kpc galactocentric radius in the plane of the sky (we
derive 11.4 kpc from the coordinates given in their paper plus our
assumed distance), or 35.1 kpc galactocentric radius on the
extrapolated M31 disk. If this is disk, then finding non-ancient stars
seems perfectly normal, and we may wonder why there are no
{\em younger} stars present. Perhaps this is evidence that star formation in
the outer M31 disk was quenched after about half a Hubble time. 
{\em Preliminary} Keck/DEIMOS velocities of
stars around the the \citet{brown} field indicate
a velocity dispersion of 85 km s$^{-1}$ centered on the systemic
velocity of M31 (J. S. Kalirai (2005), private communication). Compared to
the 156 km s$^{-1}$ line-of-sight velocity dispersion of the M31
globular clusters, this is neither a dynamically cold
thin disk nor halo, but probably something akin to the Milky Way's
thick disk component and very much in line with our speculation of a
dynamically heated ancestral disk. Recently, \citet{chap} report
disklike kinematics between 15 and 40 kpc, but with a much smaller
velocity dispersion (30 km s$^{-1}$).

By counting red giants in the vicinity of M31, \citet{ferg} 
established the presence of dynamical clumping or streaming at radii
out to 55 kpc in an elliptical survey region shaped like an
extrapolation of the disk. Some substructure detected strongly
resembles merging events, but some of the features inside $\sim$30 kpc
also look like flocculent spiral structure. Companions M32 and NGC 205
project on the sky {\em inside} these structures. The portions of the
\citet{ferg} Fig. 2 map outside $\sim$30 kpc look rather uniform
(except for the substructures) but this may be due to Galactic
foreground stars that were not subtracted. The \citet{ferg} maps
certainly do not rule out that the outer parts of M31 are dominated by
a disturbed and tidally shredded disk.

Our synthesis of these considerations is that M31 formed a gaseous,
well-mixed disk and a Milky Way-like halo in the first half of a
Hubble time. This ancestral disk need not be oriented exactly the same
way as the present-day disk. Chemical enrichment proceeded in a
``disky'' way via rotational mixing, galactic fountains, and the
occasional supernova wind. Over time, the gas supply ran short so the
outer disk stopped forming stars. Additionally, late accreting
satellite galaxies made their dynamical presence felt, adding stellar
streams and tidally disturbing the disk stars that were already
present. The advent of more measurements of stellar kinematics in the
outer disk of M31 should reveal the presence of streams from disrupted
satellites and also the structure of the postulated disturbed
ancestral disk that may have a variable velocity dispersion as a
function of galactocentric radius or even gaps and warps caused by
strong perturbers. 

In particular, if M32 has been orbiting M31 for more than a few orbits
it will certainly be well on the way to clearing a gap in the
ancestral stellar disk due to tidal resonance effects. Finding the
size and shape of such a gap could be used to constrain M32's orbit and
infer how long it has been a satellite of M31.

\acknowledgements 

We thank P. Guhathakurta and P. I. Choi for providing M31 image mosaic
data. P. B. Stetson provided both the DAOPHOT program and valuable
advice. S. Oey provided feedback of inestimable worth. S. C. and
L. A. M. acknowledge support from the National Science and Engineering
Research Council of Canada. We thank J. S. Kalirai for permission to
publish his preliminary value for the velocity dispersion at $R_d=35$
kpc, and J. L. Serven for coining the term ``Worthey Gap,'' although
we are not sure if it applies to M31 or G. W.  This work was supported
by grant AR-08745.01-A from Space Telescope Science Institute.

\begin{deluxetable}{lccccccccc}
\rotate
\tablecolumns{10}
\tablewidth{0pt}
\tablecaption{Pointings}
\tablehead{
\colhead{Internal} & \colhead{Filters} & \colhead{$a_{\rm isophote}$} &
\colhead{ $R_d$\tablenotemark{b} } & \colhead{ $R_h$\tablenotemark{c} } &
\colhead{Original HST }   & \colhead{Weight}    & 
\colhead{Number of } & \colhead{$<$[M/H]$>$} & \colhead{FWHM} \cr 
\colhead{Label} & \colhead{} & \colhead{(Arcseconds)} & \colhead{(kpc)} & \colhead{(kpc)} & 
\colhead{Proposal} & \colhead{} & \colhead{Stars Counted} &\colhead{} &\colhead{} 
}

\startdata

Out03 & F555,F814 & 8970 & 33.5 &33.3 & 5464 & 0.02860\tablenotemark{a} & 158 & $-0.43$ & 0.57\tablenotemark{a} \\
Out04 & F555,F814 & 13295& 49.6 &18.9 & 5112 & 0.02860\tablenotemark{a} & 215 & $-0.46$ & 0.57\tablenotemark{a} \\
Out05 & F555,F814 & 8520 & 30.8 &19.0 & 5420 & 0.02860\tablenotemark{a} & 189 & $-0.46$ & 0.57\tablenotemark{a} \\
In01 & F555,F814  & 7270 & 27.1 &23.2 & 6859 & 0.03012 & 356 & $-0.40$ & 0.65 \\
In02 & F555,F814  & 5685 & 21.2 &8.8 & 6734 & 0.05449 & 1008 & $-0.50$ & 0.68 \\
In03 & F555,F814  & 4215 & 15.7 &14.5 & 6671 & 0.09507 & 1336 & $-0.56$ & 0.60 \\
In04 & F606,F814  & 2940 & 11.0 &3.0 & 6664 & 0.10487 & 529 & $-0.29$ & 0.63 \\
In05 & F555,F814  & 3510 & 13.1 &4.4 & 5322 & 0.06051 & 323 & $-0.46$ & 0.58 \\
In06 & F555,F814  & 3090 & 11.5 &8.3 & 6859 & 0.05077 & 371 & $-0.18$ & 0.58 \\
In07 & F606,F814  & 1050 & 3.9 &2.6 & 5971 & 0.43257 & 550 & $-0.11$ & 0.63 \\
In11 & F555,F814  & 1530 & 5.7 &3.8 & 6431 & 0.14300 & 631 & $-0.24$ & 0.58 \\
\enddata

\tablenotetext{a}{All three ``out'' fields were summed before being assigned 
this quantity.}
\tablenotetext{b}{$R_d$ is the galactocentric radius assuming a circular
inclined disk geometry and a distance of 0.77 Mpc.}
\tablenotetext{c}{$R_h$ is the galactocentric radius in the plane of the sky, that is, appropriate for a spherical geometry.}
\label{tab1}
\end{deluxetable}


\begin{deluxetable}{rrrrrrrrrrr}
\tablecolumns{11}
\tablewidth{0pt}
\tablecaption{Abundance Distributions}
\tablehead{
\colhead{[M/H]} & \colhead{Out} & \colhead{In01} &
\colhead{In02} & 
\colhead{In03}   & \colhead{In04}    & 
\colhead{In05} & \colhead{In06} & \colhead{In07} & \colhead{In11} 
& \colhead{Sum}  }
\startdata

 -2.15 & 0.000 & 0.005 & 0.005 & 0.000 & 0.000 & 0.000 & 0.000 & 0.000 & 0.003 & 0.001 \\
 -2.05 & 0.000 & 0.004 & 0.004 & 0.000 & 0.003 & 0.000 & 0.000 & 0.002 & 0.003 & 0.001 \\
 -1.95 & 0.000 & 0.002 & 0.006 & 0.000 & 0.006 & 0.000 & 0.000 & 0.000 & 0.003 & 0.001 \\
 -1.85 & 0.000 & 0.001 & 0.006 & 0.000 & 0.009 & 0.000 & 0.000 & 0.000 & 0.003 & 0.001 \\
 -1.75 & 0.000 & 0.002 & 0.005 & 0.002 & 0.009 & 0.000 & 0.000 & 0.000 & 0.003 & 0.001 \\
 -1.65 & 0.000 & 0.002 & 0.006 & 0.002 & 0.006 & 0.000 & 0.000 & 0.000 & 0.003 & 0.001 \\
 -1.55 & 0.004 & 0.002 & 0.002 & 0.000 & 0.003 & 0.003 & 0.000 & 0.003 & 0.002 & 0.001 \\
 -1.45 & 0.000 & 0.007 & 0.005 & 0.000 & 0.009 & 0.003 & 0.004 & 0.003 & 0.007 & 0.004 \\
 -1.35 & 0.003 & 0.003 & 0.004 & 0.000 & 0.021 & 0.000 & 0.004 & 0.000 & 0.007 & 0.004 \\
 -1.25 & 0.003 & 0.004 & 0.010 & 0.000 & 0.015 & 0.000 & 0.000 & 0.002 & 0.008 & 0.003 \\
 -1.15 & 0.005 & 0.017 & 0.021 & 0.004 & 0.009 & 0.003 & 0.002 & 0.012 & 0.014 & 0.007 \\
 -1.05 & 0.010 & 0.014 & 0.025 & 0.006 & 0.013 & 0.006 & 0.002 & 0.010 & 0.016 & 0.008 \\
 -0.95 & 0.015 & 0.023 & 0.039 & 0.014 & 0.020 & 0.015 & 0.002 & 0.011 & 0.023 & 0.012 \\
 -0.85 & 0.023 & 0.035 & 0.051 & 0.019 & 0.027 & 0.000 & 0.002 & 0.016 & 0.048 & 0.016 \\
 -0.75 & 0.046 & 0.085 & 0.067 & 0.044 & 0.037 & 0.022 & 0.009 & 0.028 & 0.054 & 0.030 \\
 -0.65 & 0.096 & 0.107 & 0.119 & 0.054 & 0.064 & 0.019 & 0.017 & 0.049 & 0.092 & 0.047 \\
 -0.55 & 0.105 & 0.116 & 0.118 & 0.069 & 0.082 & 0.034 & 0.030 & 0.046 & 0.134 & 0.058 \\
 -0.45 & 0.119 & 0.143 & 0.124 & 0.087 & 0.124 & 0.068 & 0.072 & 0.081 & 0.120 & 0.089 \\
 -0.35 & 0.179 & 0.152 & 0.128 & 0.118 & 0.133 & 0.124 & 0.092 & 0.111 & 0.141 & 0.112 \\
 -0.25 & 0.180 & 0.119 & 0.099 & 0.151 & 0.133 & 0.121 & 0.098 & 0.111 & 0.123 & 0.113 \\
 -0.15 & 0.146 & 0.093 & 0.078 & 0.136 & 0.130 & 0.135 & 0.123 & 0.122 & 0.074 & 0.119 \\
 -0.05 & 0.047 & 0.047 & 0.052 & 0.154 & 0.094 & 0.117 & 0.144 & 0.112 & 0.049 & 0.116 \\
  0.05 & 0.019 & 0.014 & 0.021 & 0.117 & 0.041 & 0.200 & 0.196 & 0.168 & 0.045 & 0.138 \\
  0.15 & 0.002 & 0.004 & 0.003 & 0.018 & 0.010 & 0.089 & 0.110 & 0.077 & 0.010 & 0.066 \\
  0.25 & 0.000 & 0.001 & 0.001 & 0.004 & 0.002 & 0.029 & 0.052 & 0.026 & 0.008 & 0.029 \\
  0.35 & 0.000 & 0.000 & 0.000 & 0.001 & 0.000 & 0.006 & 0.026 & 0.008 & 0.004 & 0.013 \\
  0.45 & 0.000 & 0.001 & 0.000 & 0.000 & 0.002 & 0.006 & 0.014 & 0.004 & 0.004 & 0.007 \\

\enddata
\label{tab2}
\end{deluxetable}

\begin{thebibliography}{}

\bibitem[Audouze \& Tinsley(1976)]{at76} Audouze, J., \& Tinsley,
B. M. 1976, \araa, 14, 43
\bibitem[Binney \& Tremaine(1987)]{bt} Binney, J., \& Tremaine,
S. 1987, Galactic Dynamics, (Princeton University Press: Princeton)
\bibitem[Blair et al.(1982)]{blair} Blair, W. P., Kirshner, R. P.,
\& Chevalier, R. A. 1982, \apj, 254, 50
\bibitem[Bresolin, Kennicutt, \& Garnett(1999)]{bresolin} Bresolin,
F., Kennicutt, R. C., Jr., \& Garnett, D. R. 1999, \apj, 510, 104
\bibitem[Brown et al.(2003)]{brown} Brown, T. M., Ferguson, H. C.,
Smith, E., Kimble, R. A., Sweigart, A. V., Renzini, A., Rich, R. M.,
\& VandenBerg, D. A. 2003, \apj, 592, L17
\bibitem[Burkholder et al.(2001)]{burk01} Burkholder, V., Impey, C.,
\& Sprayberry, D. 2001, \aj, 122, 2318
\bibitem[Cen(2003a)]{cena} Cen, R. 2003, \apj, 591, 12
\bibitem[Cen(2003b)]{cenb} Cen, R. 2003, \apj, 591, L5
\bibitem[Chabrier(2003)]{chabrier} Chabrier, G. 2003, \pasp, 115, 763 
\bibitem[Chapman et al.(2005)]{chap} Chapman, S. C., Ibata, R.,
Ferguson, A., Irwin, M., Lewis, G., \& Tanvir, N. 2005, \baas, 37, 21.01
\bibitem[Chen et al.(2001)]{chen} Chen, B., Stoughton, C., Smith,
A., Uomoto, A., Pier, J. R., Yanny, B., Ivezic, Z., York, D. G.,
Anderson, J. E., Annis, J., Brinkman, J., Csabal, I., Fukugita, M.,
Hindsley, R., Lupton, R., \& Munn, J. A. 2001, \apj, 553, 184
\bibitem[Chiappini et al.(2000)]{chia00} Chiappini, C., Matteucci, F.,
\& Padoan, P. 2000, \apj, 528, 711
\bibitem[Chiosi et al.(1998)]{chiosi98} Chiosi, C., Bressan, A.,
Portinari, L., \& Tantalo, R. 1998, \aap, 339, 355 
\bibitem[Choi et al.(2002)]{choi02} Choi, P. I., Guhathakurta, P., \&
Johnston, K. V. 2002, \aj, 124, 310
\bibitem[Courteau(1996)]{1996ApJS..103..363C} Courteau, S.\ 1996, \apjs, 
103, 363 
\bibitem[Dennefeld \& Kunth(1981)]{denne} Dennefeld, M., \& Kunth,
D. 1981, \aj, 86, 989
\bibitem[Durrell, Harris, \& Pritchet(2001)]{durrell} Durrell, P. R., Harris, W. E., \& Pritchet, C. J. 2004, \aj, 121, 2557
\bibitem[Durrell, Harris, \& Pritchet(2004)]{dur04} Durrell, P. R., Harris, W. E., \& Pritchet, C. J. 2004, \aj, 128, 260
\bibitem[Evans et al.(2000)]{2000ApJ...540L...9E} Evans, N.~W., Wilkinson, 
M.~I., Guhathakurta, P., Grebel, E.~K., \& Vogt, S.~S.\ 2000, \apjl, 540, 
L9 
\bibitem[Ferguson et al.(2002)]{ferg} Ferguson, A. M. N., Irwin,
M. J., Ibata, R. A., Lewis, G. F., \& Tanvir, N. R. 2002, \aj, 124, 1452
\bibitem[Freedman \& Madore(1990)]{freedman} Freedman, W. L., \&
Madore, B. F. 1990, \apj, 365, 186
\bibitem[Gottesman, Hunter, \& Boonyasait(2002)]{2002MNRAS.337...34G} 
Gottesman, S.~T., Hunter, J.~H., \& Boonyasait, V.\ 2002, \mnras, 337, 34 
\bibitem[Governato et al.(2004)]{gov04} Governato, F., Mayer, L.,
Wadsley, J., Gardner, J. P., Willman, B., Hayashi, E., Quinn, T.,
Stadel, J., \& Lake, G. 2004, \apj, 607, 688
\bibitem[Grillmair et al.(1996)]{grill} Grillmair, C. J., Lauer,
T. R., Worthey, G., Faber, S. M., Freedman, W. L., Madore, B. F.,
Ajhar, E. A., Baum, W. A., Holtzman, J. A., Lynds, C. R., O'Neil,
E. J., Jr., \& Stetson, P. B. 1996, \aj, 112, 1975
\bibitem[Haywood(2001)]{hay01} Haywood, M. 2001, \mnras, 325, 1365
\bibitem[Harris \& Harris(2001)]{hh01} Harris, W. E., \& Harris,
G. L. H. 2001, \aj, 122, 3065
\bibitem[Harris \& Harris(2002)]{hh02} Harris, W. E., \& Harris,
G. L. H. 2002, \aj, 123, 3108
\bibitem[Hodge \& Kennicutt(1982)]{hk} Hodge, P. W., \& Kennicutt,
R. C. 1982, \aj, 87, 264
\bibitem[Holtzman et al.(1995)]{holtz} Holtzman, J. A., Burrows,
C. J., Casertano, S., Hester, J. J., Trauger, J. T., Watson, A. M., \&
Worthey, G. 1995, \pasp, 107, 1065
\bibitem[Huchra(1993)]{huchra} Huchra, J. P., in The Globular
Cluster-Galaxy Connection (ASP Conference Series Vol. 48),
ed. G. H. Smith and J. P. Brodie, San Francisco: Astron. Soc. Pac., 420
\bibitem[Hurley-Keller et al.(2004)]{hurley} Hurley-Keller, D.,
Morrison, H. L., Harding, P., \& Jacoby, G. 2004, \apj, accepted;
astro/ph 0408137
\bibitem[J\o rgensen(2000)]{jorgensen} J\o rgensen, B. R. 2000, \aap,
363, 947
\bibitem[Kotoneva et al.(2002)]{koto02} Kotoneva, E., Flynn, c.,
Chiappini, C., \& Matteucci, F. 2002, \mnras, 336, 879
\bibitem[Larson(1998)]{larson} Larson, R. B. 1998, \mnras, 301, 569 
\bibitem[MacArthur, Courteau, \& Holtzman (2003)]{mac} MacArthur,
L. A., Courteau, S., \& Holtzman, J. A. 2003, \apj, 582, 689
\bibitem[Malinie et al.(1993)]{malinie93} Malinie, G., Hartmann,
D. H., Clayton, D. D., \& Mathews, G. J. 1993, \apj, 413, 633 
\bibitem[Morrison et al.(2004)]{morrison} Morrison, H. L., Harding,
P., Perrett, K., \& Hurley-Keller, D. 2004, \apj, 603, 87
\bibitem[Oey(2000)]{oey00} Oey, M. S., 2000, \apjl, 542, L25
\bibitem[Padoan et al.(1997)]{padoan97} Padoan, P., Jiminez, R., \&
Antoniccio-Delogu, V. 1997, \mnras, 288, 145
\bibitem[Perrett et al.(2002)]{perrett} Perrett, K. M., Bridges,
T. J., Hanes, D. A., Irwin, M. J., Brodie, J. P., Carter, D., Huchra,
J. P., \& Watson, F. G. 2002, \aj, 123, 2490
\bibitem[Racine(1991)]{racine} Racine, R. 1991, \aj, 101, 865
\bibitem[Reed et al.(1994)]{reed94} Reed, L. G., Harris, G. L. H.,
\& Harris, W. E. 1994, \aj, 107, 555
\bibitem[Reitzel \& Guhathakurta(2002)]{rg02} Reitzel, D. B., \&
Guhathakurta, P. 2002, \aj, 124, 234
\bibitem[Rejkuba et al.(2004)]{rej04} Rejkuba, M., Greggio, L.,
Harris, W. E., Harris, G. L. H., \& Peng, E. W. 2004, \baas, 205, 92.15
\bibitem[Rejkuba et al.(2005)]{rej05} Rejkuba, M., Greggio, L.,
Harris, W. E., Harris, G. L. H., \& Peng, E. W. 2005, \apj, submitted
\bibitem[Renda et al.(2005)]{renda} Renda, A., Daisuke, K., Fenner,
Y., \& Gibson, B. K. 2005, \mnras, 356, 1071
\bibitem[Rich, Mighell, \& Neill(1996)]{rich} Rich, R. M., Mighell,
K. J., \& Neill, J. D. 1996, in Formation of the Galactic Halo, Inside
and Out (ASP Conference Series, Vol 92) ed. H. Morrison \&
A. Sarajedini, 544
\bibitem[Sandage(1987)]{sandage} Sandage, A. 1987, \aj, 93, 610
\bibitem[Scannapieco, Schneider, \& Ferrara(2003)]{scanna}
Scannapieco, E., Schneider, R., \& Ferrara, A. 2003, \apj, 589, 35
\bibitem[Schlegel, Finkbeiner, \& Davis(1998)]{schlegel} Schlegel,
D. J., Finkbeiner, D. P., \& Davis, M. 1998, \apj, 500, 525
\bibitem[Schneider et al.(2002)]{schneider} Schneider, R., Ferrara,
A., Natarajan, P., \& Omukai, K. 2002, \apj, 571, 30
\bibitem[Searle(1977)]{searle77} Searle, L., in The Evolution of
Galaxies and Stellar Populations, ed. B. M. Tinsley \& R. B. Larson
(New Haven: Yale Univ. Obs.), 219
\bibitem[Searle \& Sargent(1972)]{ss72} Searle, L., \& Sargent,
W. L. W. 1972, \apj, 173, 25
\bibitem[Sil'chenko et al.(1998)]{silch} Sil'chenko, O. K.,
Burenkov, A. N., \& Vlasyuk, V. V. 1998, \aap, 337, 349
\bibitem[Smail et al.(2002)]{smail} Smail, I., Owen, F. N., Morrison,
G. E., Keel, W. C., Ivison, R. J., \& Ledlow, M. J. 2002, \apj, 581, 844
\bibitem[Stanek \& Garnavich(1998)]{stan} Stanek, K. Z., \&
Garnavich, P. M. 1998, \apj, 503, L131
\bibitem[Stetson(1987)]{stet} Stetson, P. B. 1987, \pasp, 99, 191
\bibitem[Tinsley(1975)]{tin75}Tinsley, B. M. 1975, \apj, 197, 159
\bibitem[Tosi(1988)]{tosi} Tosi, M. 1988, \aap, 197, 33
\bibitem[van den Bergh(2000)]{vdb} van den Bergh, S. 2000, The
Galaxies of the Local Group (Cambridge: Cambridge University Press)
\bibitem[van den Bergh(1969)]{vdb69} van den Bergh, S. 1969, ApJS, 19, 145
\bibitem[van der Kruit(1989)]{vdk} van der Kruit, P. C. 1989, in The
Milky Way as a Galaxy, ed. G. Gilmore, I. R. King, and P. C. van der
Kruit (Sauverny: Geneva Observatory), 331
\bibitem[Kroupa \& Weidner(2003)]{weid03} Kroupa, P., \& Weidner,
C. 2003, \apj, 598, 1076
\bibitem[Walterbos \& Kennicutt(1987)]{wk87} Walterbos, R. A. M., \&
Kennicutt, R. C., Jr. 1987 \aaps, 69, 311
\bibitem[Whitmore et al.(1999)]{whitmore} Whitmore, B., Heyer, I.,
\& Casertano, S. 1999, \pasp, 111, 1559
\bibitem[Wise \& Gilmore(1995)]{wyse95} Wyse, R. F. G., \& Gilmore,
G. 1995, \aj, 110, 2771
\bibitem[Worthey(1994)]{w94} Worthey, G. 1994, \apj, 95, 107
\bibitem[Worthey, Dorman, \& Jones (1996)]{wdj} Worthey, G., Dorman,
B., \& Jones, L. A. 1996, \aj, 112, 948
\bibitem[Worthey(1998)]{w98} Worthey, G. 1998, \pasp, 110, 888
\bibitem[Worthey et al.(2004)]{w04} Worthey, G., Mateo, M.,
Alonso-Garc\'{\i}a, J., \& Espa\~na, A. L. 2004, \pasp, 116, 295 
\bibitem[Wyse \& Gilmore(1995)]{wyse} Wyse, R. F. G., \& Gilmore,
G. 1995, \aj, 110, 2771

\end{thebibliography}
\end{document}